\documentclass[lettersize,journal]{IEEEtran}
\usepackage{amsmath,amsfonts}
\usepackage{algorithmic}
\usepackage{algorithm}
\usepackage{array}
\usepackage[caption=false,font=normalsize,labelfont=sf,textfont=sf]{subfig}
\usepackage{textcomp}
\usepackage{stfloats}
\usepackage{url}
\usepackage{verbatim}
\usepackage{graphicx}
\usepackage{cite}
\usepackage{balance}
\usepackage{xcolor}
\usepackage{colortbl}
\usepackage{color}
\usepackage{soul}

\usepackage{array}
\newcounter{tcbcounter}
\usepackage{booktabs}
\usepackage{multirow}
\usepackage{tikz}
\usepackage{tabularx}
\usepackage{longtable}
\usepackage{enumitem}
\usepackage[most]{tcolorbox}
\usepackage{amssymb}      % For checkmarks and xmarks
\usepackage{pifont}  
\usepackage{graphicx}
\usepackage{svg}

\usepackage{algorithmic}
\usepackage{graphicx}
\usepackage{textcomp}
\usepackage{comment}
\usepackage{hyperref}
\usepackage{pifont}
\usepackage{fontawesome5}

\usepackage{listings}
\usepackage[utf8]{inputenc}

\definecolor{lightgray}{gray}{0.95}
\lstdefinelanguage{json}{
    basicstyle=\ttfamily\small,
    numbers=left,
    numberstyle=\tiny,
    stepnumber=1,
    numbersep=8pt,
    showstringspaces=false,
    breaklines=true,
    backgroundcolor=\color{lightgray},
    string=[s]{"}{"},
    morecomment=[l]{//},
    commentstyle=\color{gray},
    keywordstyle=\color{blue},
}

\usepackage[textsize=scriptsize]{todonotes}

\tcbuselibrary{skins}
\NewTotalTColorBox[auto counter]{\Definition}{ +m }{ 
    breakable,
    notitle,
    colback=blue!3!white,
    frame hidden,
    boxrule=0pt,
    enhanced,
    sharp corners,
    borderline west={4pt}{0pt}{blue!20!white},
     boxsep=2pt,  % Decrease padding around the content
    left=7pt,    % Reduce left inner padding
    right=2pt,   % Reduce right inner padding
    top=2pt,     % Reduce top inner padding
    bottom=2pt,  % Reduce bottom inner padding
}{
    \textcolor{green!50!black}{
        \sffamily
    }%
    #1
}

\NewTotalTColorBox[auto counter]{\Def}{ +m }{ 
    breakable,
    notitle,
    colback=blue!3!white,
    frame hidden,
    boxrule=0pt,
    enhanced,
    sharp corners,
    borderline west={4pt}{0pt}{blue!20!white},
     boxsep=2pt,  % Decrease padding around the content
    left=7pt,    % Reduce left inner padding
    right=2pt,   % Reduce right inner padding
    top=2pt,     % Reduce top inner padding
    bottom=2pt,  % Reduce bottom inner padding
}{
    \refstepcounter{tcbcounter}%
    \textcolor{black}{
        \sffamily
        \textit{Def~\thetcbcounter.}%
    }%
    #1
}

\newcommand{\cmark}{\cellcolor{green!15}\textcolor{green!60!black}{\checkmark}}
\newcommand{\xmark}{\cellcolor{red!15}\textcolor{red!70!black}{\ding{55}}}

\newlength\MAX
\begin{document}

\title{GoLeash: Mitigating Golang Software Supply Chain Attacks with Runtime Policy Enforcement}

\author{
\IEEEauthorblockN{
Carmine Cesarano\IEEEauthorrefmark{1},
Martin Monperrus\IEEEauthorrefmark{1},
Roberto Natella\IEEEauthorrefmark{2}
}\\
\IEEEauthorblockA{
\IEEEauthorrefmark{1}KTH Royal Institute of Technology, Sweden
}\\
\IEEEauthorblockA{
\IEEEauthorrefmark{2}Gran Sasso Science Institute, Italy
}
}

\markboth{Journal of \LaTeX\ Class Files,~Vol.~14, No.~8, August~2021}%
{Shell \MakeLowercase{\textit{et al.}}: A Sample Article Using IEEEtran.cls for IEEE Journals}


\maketitle

\begin{abstract}
Modern software supply chain attacks consist of introducing new, malicious capabilities into trusted third-party software components, in order to propagate to a victim through a package dependency chain. 
These attacks are especially concerning for the Go language ecosystem, which is extensively used in critical cloud infrastructures. 
We present GoLeash, a novel system that applies the principle of least privilege at the package-level granularity, by enforcing distinct security policies for each package in the supply chain. This finer granularity enables GoLeash to detect malicious packages more precisely than traditional sandboxing that handles security policies at process- or container-level. Moreover, GoLeash remains effective under obfuscation, can overcome the limitations of static analysis, and incurs acceptable runtime overhead.
\end{abstract}


\section{Introduction}

In recent years, software supply chain (SSC) attacks have jumped to the forefront of cybersecurity concerns. Rather than attacking end-users directly, adversaries now infiltrate the software development pipeline. Attackers have been injecting harmful code into widely-used software packages through social engineering, malicious commits, typosquatting, and hijacking unmaintained repositories \cite{cisa_ssc, cox2025fifty}. 
Once a compromised package is released, downstream projects are infected by malicious code through normal dependency resolution and build processes. Their users then become susceptible to credential theft, resource hijacking, and remote command execution \cite{eventstream_attack}. 

These supply chain attacks are especially alarming for the Go language (Golang) ecosystem. This language has been heavily used for critical cloud software \cite{go_cacm}, including Kubernetes \cite{kubernetes}, Docker \cite{docker}, Terraform \cite{terraform}, Etcd \cite{etcd}, and several others mission-critical infrastructures \cite{go_case_studies}. Injecting malicious code into these Golang software projects would enable unauthorized control over cloud infrastructures, with potentially devastating consequences. 
We are already witnessing supply chain attacks against the Golang ecosystem, including the recent typosquatting attacks to the BoltDB database \cite{go_supply_chain_attack_boltdb} and Hypert testing library \cite{hypert}, and the huge number of Go packages found on GitHub that are vulnerable to repojacking \cite{repojacking_go}. In this paper, we aim to improve the state of the art of mitigating software supply chain attacks on Golang critical applications.

We present \emph{GoLeash}, a novel approach and its research prototype for mitigating malicious behavior of Go packages once they have been compromised. 
GoLeash is inspired by the ``principle of least privilege'', that is, the idea that accesses should be limited to the minimal set that is feasible and practical \cite{saltzer1975protection,smith2012contemporary}. 
GoLeash has two modes: 1) an analysis mode to automatically infer the policies of Go packages; and 2) an enforcement mode that restricts the behavior of Go packages at run-time, by checking that Go packages comply with the package policy. 
Go packages in the supply chain are forbidden to perform operations with \emph{capabilities} that they are not supposed to have (for example, opening a network connection in a package for local data processing). 

% how it works
GoLeash monitors Go programs at run-time, using the eBPF framework \cite{ebpf_foundation}, to collect information about capabilities accessed by Go packages and automatically build a security policy (analysis). Thus, it enforces Go packages to have the minimal set of capabilities required to perform their intended purpose (enforcement).

% novelty claim
GoLeash is a novel approach for policy enforcement. 
Existing approaches for monitoring and sandboxing of system calls for Go~\cite{seccomp_linux, gvisor} only apply to entire processes and containers, which makes enforcement too coarse-grained to mitigate supply chain attacks. If a package in the program is allowed to use a capability (e.g., a privileged system call), the entire process or container needs to be assigned that capability. 
GoLeash works at a finer grain, since it is able to trace back which Go package is using a capability at a given moment, thus it can enforce separate policies across different Go packages. 
While runtime sandboxing is itself a well-studied area, the contribution of GoLeash lies doing per-dependency sandboxing with three contributions: per-package syscall attribution within a single statically-linked binary; automated per-package policy inference from observed execution; and an enforcement design that holds under a threat model in which dependencies themselves are treated as adversarial and untrusted. As we demonstrate in our experiments, this makes GoLeash more precise and less prone to omissions than alternatives.

% evaluation
To evaluate GoLeash, we systematically inject malicious behaviors (e.g., exfiltrate or infiltrate files) into the software supply chain of real-world, complex applications such as kubernetes. Our evaluation shows that GoLeash is able to successfully identify malicious packages in $98\%$ of cases, compared to $32\%$ in the case of coarse-grained monitoring. Moreover, our approach is effective even when obfuscation techniques are used for hiding the malicious code. GoLeash has an acceptable overhead, by introducing $1.3$--$5.1\%$ per-request CPU overhead.

In summary, the main contributions of this work are:
\begin{itemize}[leftmargin=1.2em, itemsep=0pt]
    \item A new approach for run-time capability enforcement based on tracing and enforcing specific capabilities for individual packages in a software supply chain. 
    \item GoLeash, an efficient implementation of the approach for Golang, based on the eBPF observability framework, available as open-source for future research \cite{goleash_artifact}. GoLeash is able to both learn and enforce package-level dependency policies. 
    \item An evaluation of GoLeash on five real-world, complex Go projects, incl. one of the largest Go application ever, Kubernetes, in terms of software supply chain attack mitigation and performance overhead.
\end{itemize}

The paper is structured as follows. In Section~\ref{sec:background_motivation}, we provide technical background and motivation for the paper. In Section~\ref{sec:threat_model}, we present the threat model and assumptions. Section~\ref{sec:design} presents the design of GoLeash. Section~\ref{sec:evaluation} evaluates the approach. Section~\ref{sec:related} analyzes related work. Section~\ref{sec:discussion} discusses use cases and limitations of GoLeash. Section~\ref{sec:conclusion} concludes the paper.

% -------------------------------------------------------------------------------------------%

\section{Background}
\label{sec:background_motivation}

\subsection{Real-world Go Supply Chain Attack}
\label{subsec:motivating}
A concrete real-world example of the pressing supply chain threats in Go is the recent compromise of the popular \texttt{boltdb/bolt} module \cite{go_supply_chain_attack_boltdb}. An adversary published a typosquatted module impersonating the original library. While appearing entirely benign and preserving its normal database functionality, the malicious module stealthily introduced remote code execution (RCE) capabilities at run-time. Specifically, the compromised package established a persistent, obfuscated network connection to a command-and-control (C2) server using Go's standard \texttt{net.Dial} function. Once connected, the package awaited commands from attackers and executed arbitrary shell instructions through the \texttt{os/exec.Command} API, effectively enabling unrestricted file manipulation, process control, and network access, which are capabilities never intended by the original BoltDB library. After the malicious package was cached by the Go Module Mirror, the attacker rewrote the Git tag to point to a clean commit, concealing the backdoor in the public repository. However, the proxy had already cached the malicious pseudo-version, which downstream applications continued to fetch long after the repository appeared clean. This allowed the backdoor to persist in the Go ecosystem for years.

Software supply chain attacks are powerful.
Attackers increasingly leverage sophisticated obfuscation techniques, ranging from encryption of malicious payloads to more subtle strategies. Go's runtime provides uniquely powerful features such as CGO (integration of C code into Go binaries), inline assembly, dynamically loaded plugins, external binary execution, and reflection-based dynamic function invocation, which can all be used for malicious purposes. These techniques bypass traditional static analysis \cite{molina2025light}, and are best addressed through run-time solutions \cite{khoury2012security}. However, as also discussed in Section~\ref{sec:related}, there is no run-time solution to address software supply chain attacks for Go.

\subsection{The Go Module System}
Modern software development heavily relies on third-party code reuse, and the Go ecosystem is no exception. GitHub hosts over 1.8 million Golang modules \cite{github}. This vast ecosystem highlights the scale at which open-source third-party code is made available and actively consumed in Go development. Large-scale, industrial Go applications often depend on hundreds of external packages. A notable example is Kubernetes, which currently pulls hundreds direct and indirect Go dependencies to build \cite{depsdev}.
The Go module system \cite{go-modules-blog} provides comprehensive support for dependency management, with semantic versioning, reproducible builds, and compatibility guarantees. 

In Go, a module is a collection of packages that are released, versioned, and distributed together. Modules may be downloaded directly from version control repositories (commonly Git repositories), or from module proxy servers. Modules are downloaded and built using the standard \texttt{go} command-line tool. A package refers to a directory containing one or more Go source files in the same namespace and represents the fundamental unit of compilation and encapsulation. A module may consist of one or many packages. A project is a library or an application hosted in a repository with one or several modules.

When a Go project is built, all required modules and packages are bundled into a single statically-linked binary. The binary contains the Go runtime \cite{go-runtime-docs}, which manages initialization, garbage collection, and concurrency. In order to achieve high performance, the Go runtime is designed to be simple and lightweight.

The Go module system introduces unique technical challenges. In particular, the activity of packages inside a Go program is opaque from the point of view of monitoring tools. Since the compiled Go program is a flat binary, the OS has no visibility on which Go package is invoking a system call. Moreover, the Go runtime does not provide security features to analyze and manage accesses to platform features and APIs, such as in the Security Manager for the Java language. This lack of observability makes it easier for attackers to slip malicious code into Go programs.

\subsection{The eBPF Runtime Monitoring System}
The Extended Berkeley Packet Filter (eBPF) \cite{ebpf_foundation} is a Linux kernel technology that executes small, sandboxed programs in kernel space. eBPF offers a versatile framework for dynamic tracing and performance analysis. By attaching eBPF programs to kernel hooks, one can analyze kernel events at runtime with minimal overhead. This includes capturing information on system call IDs and arguments, and the current stack trace, a feature we will extensively use in GoLeash. eBPF is an excellent tool for security. It has been used to implement fine-grained policies and analyses on system-wide and application-specific behavior \cite{falco, cilium, tetragon, isovalent}. 
Since eBPF is embedded in the kernel, it ensures that telemetry data collection remains efficient and isolated. To sum up, eBPF is a mature solution to detect runtime anomalies and enforce run-time constraints based on security policies. 

\Definition{
Existing security solutions fall short to counter sophisticated supply chain attacks in Go. On the one hand, static analysis struggles against advanced forms of obfuscation \cite{molina2025light}.
On the other hand, current runtime monitoring solutions enforce coarse-grain policies at the process- or container-level \cite{khoury2012security}.
Fine-grain, package-level policies are required to detect and mitigate modern software supply chain attacks. 
}

\section{Threat Model}
\label{sec:threat_model}

GoLeash is designed to mitigate software supply chain attacks targeting third-party Go dependencies. Our threat model assumes attackers compromise initially trusted dependencies through malicious updates, typically via compromised developer credentials or repository hijacking \cite{siadati2024devphish}. The primary attacker objective is to inject malicious code into legitimate-looking packages to perform unauthorized operations at runtime, such as accessing and manipulating the filesystem, establishing network communications, and executing arbitrary commands. That is, third-party dependencies are considered untrusted and potentially compromised, as they can embed arbitrary code execution capabilities.

Attackers can add malicious capabilities in pure Go code, either using obfuscation or not. Attackers in our model may also employ sophisticated techniques beyond Go code \cite{cesarano2023gosurf}, using mechanisms such as CGO integration (C binding), inline assembly, dynamically loaded plugins, and execution of external binaries. Such advanced methods evade detection from  static analysis tools, as malicious behaviors only become observable during actual program execution. 

% what we don't do 1
The Go run-time environment, the operating system kernel, and the GoLeash framework itself are part of the trusted computing base. This implies that a threat actor cannot manipulate the context of system calls or compromise the integrity of Go's execution model (including function dispatch, stack trace construction, and goroutine scheduling), nor tamper with system-level components such as kernel tracepoints, eBPF programs, or GoLeash's user-space data structures. We also assume that the protected application is deployed in a form that preserves sufficient runtime metadata (e.g., Go function symbols) to enable analysis by our approach.
GoLeash's trusted base has the standard trust assumptions of widely deployed eBPF-based security tools (Falco~\cite{falco}, Tetragon~\cite{tetragon}, Cilium~\cite{cilium}): loading eBPF programs requires \texttt{CAP\_SYS\_ADMIN} (or \texttt{CAP\_BPF}), so subverting GoLeash already requires privileges that defeat any user-space defense. Known eBPF threats such as verifier bypasses (e.g., CVE-2021-3490) and stealthy persistence (e.g., BPFDoor) are properties of the subsystem rather than GoLeash, with the same applicable mitigations.

% what we don't do 2
We do not cover attacks executed during earlier phases of software development, such as attacks triggered by test functions, Go generate scripts, Makefiles, or other build and deployment tooling. These attacks primarily compromise developer machines rather than production systems.

% What we don't do 3
We do not target cases where a dependency silently introduces latent vulnerabilities, such as weakened input validation or logic bugs, without directly exercising new or unauthorized system-level capabilities at runtime. These are better addressed by static analysis and testing tools. However, if such vulnerabilities are later exploited to trigger unauthorized behaviors (e.g., by injecting commands or activating unexpected capabilities), GoLeash can detect the resulting deviation during execution.

\Definition{In this paper, our threat model is unauthorized runtime behavior in production, introduced via third-party Go dependencies compromised with malicious code.}

% -------------------------------------------------------------------------------------------%
\section{Design of GoLeash}
\label{sec:design}
In this section, we present the design requirements, key design choices, and implementation details of GoLeash.

\subsection{Design Rationales}
GoLeash is designed to meet the following requirements, which are essential for the practical adoption of the approach: 

\begin{itemize}[leftmargin=1.2em, labelsep=0.2em] 
    \item[\faLayerGroup] Infer package-level security policies without developer manual work.
    \item[\faEye] Enforce security policies without changes to source code, build pipelines, and the Go runtime. 
    \item[\faProjectDiagram] Handle real-world, complex applications, including ones involving multiple processes and binaries.
    \item[\faUserSecret] Detect malicious dependencies even when the malicious behavior is obfuscated.
\end{itemize}

Generating accurate system call policies for complex programs has long been a challenge in system security. Precise enforcement requires understanding which capabilities an application actually uses in a given deployment context, without either overestimating its needs or missing critical behaviors.
 
Static analysis is a natural candidate for policy generation, as it can inspect all reachable code paths. However, it faces two opposing limitations. On one hand, it often yields an \emph{over-approximation}, including capabilities associated with code that is imported but not exercised under real workload conditions \cite{rostamipoor_confine_2023, canella2021automating, demarinis2020sysfilter}. For example, a package may expose file or network access functions that remain unused in a particular deployment configuration, yet static analysis cannot safely discard them. This can result in coarse or overly permissive policy. On the other hand, static analysis may \emph{under-approximate} behavior due to dynamic language features. In Go, applications commonly dynamically invoke capabilities through mechanisms like function pointers, reflection, foreign function interfaces (CGO), or dynamically loaded plugins \cite{cesarano2023gosurf}. Such constructs hinder static reasoning, especially for fine-grained per-package enforcement.
 
Dynamic analysis provides a pragmatic and effective means of approximating a program's runtime capabilities. Instead of reasoning about all possible control-flow paths, it records actual execution behavior and derives policies from observed system interactions. Prior work has demonstrated its utility in capturing least-privilege profiles across various domains, including Linux containers and malware analysis, by leveraging exercised traces as sound approximation of intended program behavior \cite{comparetti2010identifying, wan2017mining, wan2019practical}.

GoLeash builds on this foundation and dynamically constructs per-package capability profiles from concrete executions. Dynamic analysis allows it to precisely attribute runtime behavior to individual packages without assuming developer annotations. As a result, it can detect when a dependency  exercises malicious system-level capabilities not observed during benign training runs. The specific mechanism for policy construction and enforcement is detailed in Section~\ref{sec:architecture}.

\subsection{Architecture}
\label{sec:architecture}
\begin{figure}[t]
  \centering
  \includegraphics[width=1\linewidth]{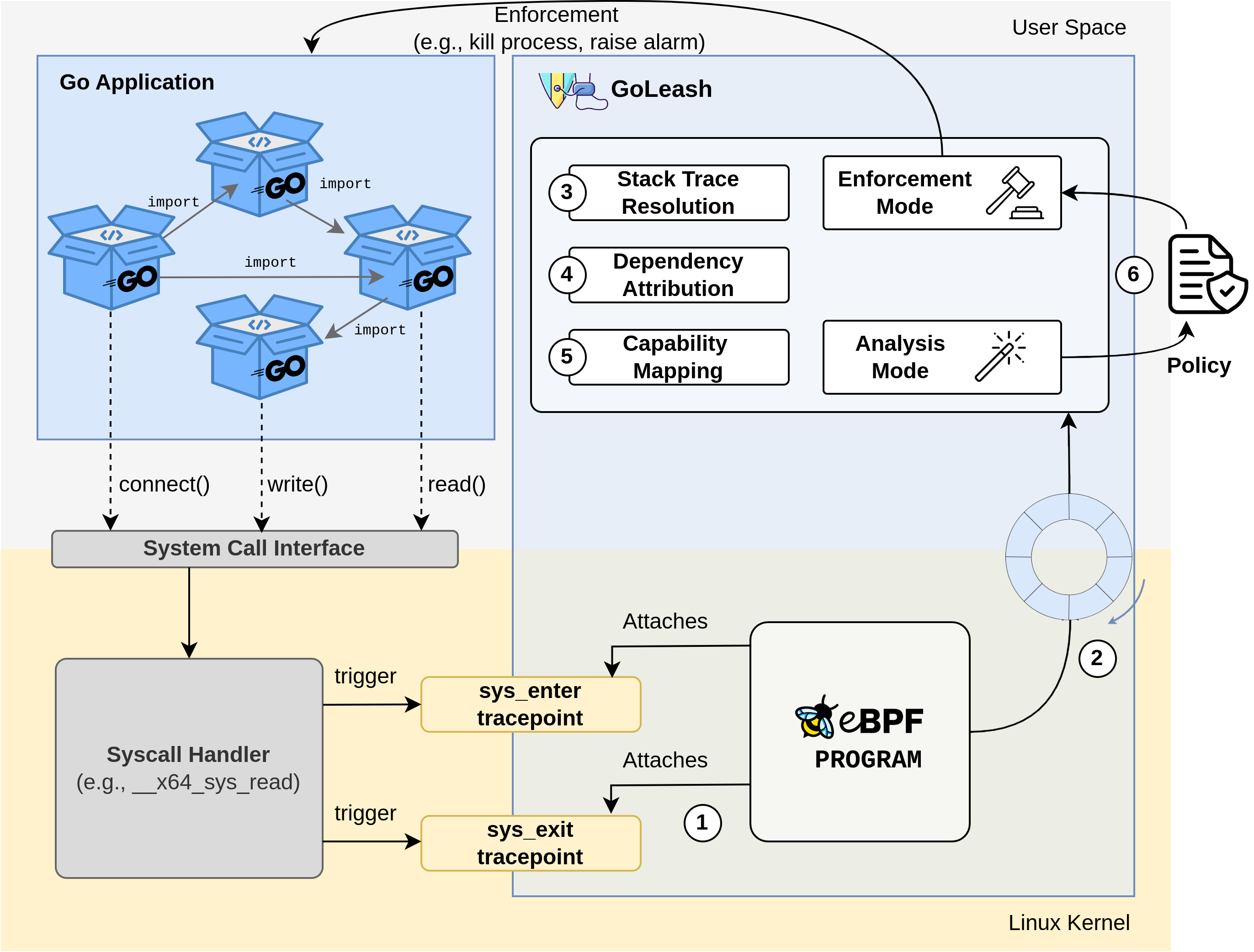}
  \caption{GoLeash Architecture}
  \label{fig:architecture_overview}
\end{figure}

Figure~\ref{fig:architecture_overview} illustrates the high-level architecture of GoLeash. The system consists of a kernel-space tracing component and a user-space analysis engine.

The workflow includes the following stages. (1) GoLeash inserts tracing probes in the kernel to capture system call (syscall) events. (2) These events are pushed into a ring buffer, and asynchronously collected in user space, where (3) the stack trace is resolved. (4) GoLeash maps the stack trace to the originating Go packages, and (5) classifies the syscall into a capability. Finally, (6) GoLeash either uses the event to build a policy (\emph{analysis mode}), or it checks whether the event matches against a policy to detect violations (\emph{enforcement mode}). In the first case, the allowlist is saved for later use. In the second case, enforcement actions are triggered. 

This architecture allows GoLeash to enforce least-privilege boundaries across dependencies, as opposed to coarse-grain process-based privileges. The system is designed to operate with no changes to build and deployment pipelines, it does not require recompilation or dedicated instrumentation of the protected application.

\subsection{System Call Tracing}
At the core of GoLeash lies a low-level monitoring infrastructure built on top of the eBPF framework. GoLeash attaches eBPF programs to syscall-related tracepoints, allowing it to intercept every system call made by the target application in real time. A Linux tracepoint provides a static hook point to call a function provided at runtime. Specifically, eBPF programs are attached to Linux kernel tracepoints to capture syscall invocations \cite{linux_tracepoints}. 

Each captured syscall event includes two key pieces of information: the \emph{syscall identifier} and a \emph{stack trace identifier}, which references the user-space call stack that led to the syscall. 

To reduce noise and ensure that only relevant syscalls are captured, GoLeash filters events based on the \emph{command name} of the traced binary, that is, the name of the executable image as reported by the kernel comm field, independently of how the binary dispatches its internal functionality. All PIDs sharing this command name are tracked.
This design supports real-world scenarios, such as multi-process Go applications, and horizontally-scaled replicas in containerized environments.

\subsection{Dependency Attribution}
After system call events are captured in kernel space, GoLeash's user-space component analyzes these events, to attribute each syscall to the specific Go package responsible for triggering it. This attribution is critical for building precise capability profiles and precise policy enforcement at the package level.

Each syscall event delivered from the kernel includes a stack trace identifier, which corresponds to a snapshot of the user-space call stack captured at the time of invocation. In user space, GoLeash resolves this identifier into a full stack trace by referencing the \texttt{stack\_id} and its associated memory snapshots. To interpret the raw return addresses in the stack trace, GoLeash maps each address to the corresponding function name and Go package import path. This symbol resolution process begins by parsing the target binary's ELF symbol table, which provides a mapping from memory addresses to symbol names in unstripped binaries.

\begin{figure}[t]
  \centering
  \includegraphics[width=1\linewidth]{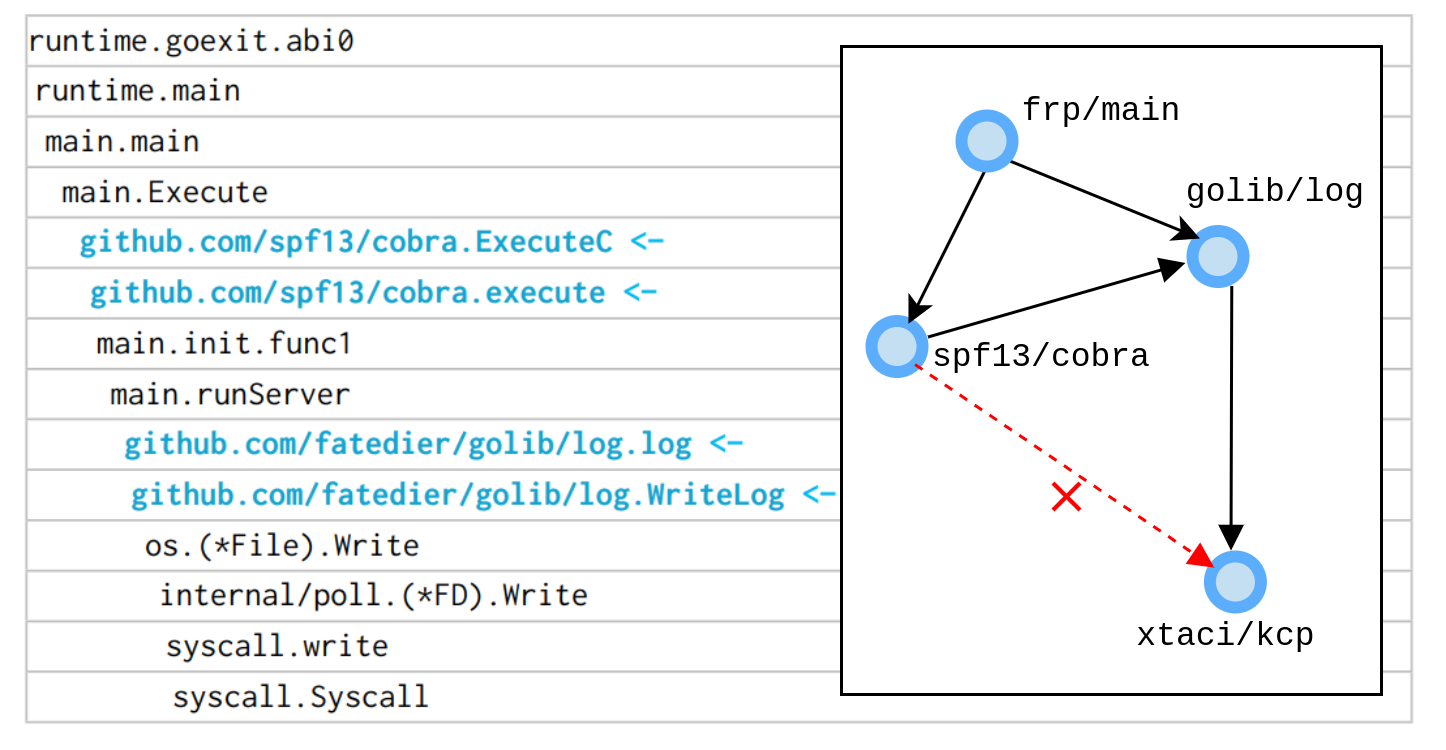}
  \caption{On the left: example stack trace with highlighted frames that are part of the call path responsible for the write syscall. On the right: dependency graph of the corresponding application being traced.}
  \label{fig:stacktrace}
\end{figure}

Once the stack trace is resolved, GoLeash traverses it from the most recent frame backwards to identify the Go package from which the syscall originates. In Go programs, system calls often pass through multiple layers of abstraction, including standard library functions (e.g., from \texttt{net/http}, \texttt{os}, and \texttt{io} \cite{go-stdlib}) and runtime helpers, before reaching the kernel. These intermediate layers are part of Go’s trusted infrastructure and are not considered responsible for the syscall. GoLeash should not misattribute behavior to trusted components. Hence, GoLeash does not assign the syscall to the topmost stack frame. Instead, it scans the full trace to find the first frame originating from an application-defined or third-party package function, and considers that frame as the initiator of the syscall. For instance, in the trace shown in Figure~\ref{fig:stacktrace} (on the left), where \texttt{os.Write} and other standard packages wrap the syscall, GoLeash skips them and correctly attributes the syscall to the frame of third-party package \texttt{fatedier/frp/server}. In addition to identifying the responsible frame, GoLeash also saves the full chain of Go packages that led to the syscall, for call path analysis discussed later in \autoref{subsec:enforcement}.

It is worth noting that the use of helper functions, dynamic dispatchers, callbacks, or interface-based function invocations is supported by GoLeash’s attribution logic. Since enforcement is grounded in the concrete runtime stack trace captured at the syscall boundary, the responsible package is always identified based on the actual dynamic invocation, regardless of the indirection mechanism used.

\subsection{Capability Mapping}
GoLeash defines a capability taxonomy (Table~\ref{tab:capability_taxonomy}) that maps low-level syscalls to semantically meaningful categories of system functionality. The taxonomy follows the design principles established in prior work on syscall-based capability analysis, in particular Capslock~\cite{capslock}. We group syscalls operating on the same kernel resource and serving the same security-relevant function, while preserving distinctions between operations with different security implications. For instance, filesystem-related syscalls are split into read and write capabilities, allowing developers to grant read access without permitting file modifications. Likewise, networking is broken down into socket creation, connection establishment, and packet transmission. 
The syscall-to-capability mapping is manually curated and verified against official Linux syscall documentation~\cite{linux_syscall_table} to ensure accuracy.
Table~\ref{tab:capability_taxonomy} reports only representative examples of syscalls for each capability category; the complete mapping of $456$ syscalls is available in our publicly available implementation of GoLeash \cite{goleash_artifact}.
We acknowledge that alternative partitions are possible; yet the chosen granularity is sufficient to express least-privilege boundaries for the supply chain attacks considered in our threat model, as confirmed by the evaluation in Section~\ref{sec:evaluation}.
During both analysis and enforcement, GoLeash uses this mapping to translate raw syscall activity into high-level capability usage.

\begin{table*}[t]
\centering
\caption{Capability Taxonomy Used in GoLeash}
\begin{tabular}{p{0.16\linewidth} p{0.20\linewidth} p{0.39\linewidth} p{0.16\linewidth}}
\toprule
\textbf{Category} & \textbf{GoLeash Capability} & \textbf{Description} & \textbf{Example Syscalls} \\
\midrule

\multirow{6}{*}{File Capabilities} 
 & \texttt{CAP\_FILE} & Manage file descriptors & \texttt{close}, \texttt{poll} \\
 & \texttt{CAP\_READ\_FILE} & Read data from files & \texttt{read}, \texttt{open}, \texttt{stat} \\
 & \texttt{CAP\_WRITE\_FILE} & Write or append data to files & \texttt{write}, \texttt{writev} \\
 & \texttt{CAP\_CREATE\_FILE} & Create new files or directories & \texttt{mkdir}, \texttt{creat} \\
 & \texttt{CAP\_DELETE\_FILE} & Remove existing files or directories & \texttt{unlink} \\
 & \texttt{CAP\_FILE\_METADATA} & Modify file metadata (permissions or ownership) & \texttt{chmod}, \texttt{chown} \\
\midrule

\multirow{4}{*}{Network Capabilities} 
 & \texttt{CAP\_CONNECT\_REMOTE} & Initiate outbound connections to remote endpoints & \texttt{socket}, \texttt{connect} \\
 & \texttt{CAP\_LISTEN\_LOCAL} & Bind to local ports to accept incoming connections & \texttt{bind}, \texttt{listen} \\
 & \texttt{CAP\_SEND\_DATA} & Transmit data over established network connections & \texttt{sendto}, \texttt{sendmsg} \\
 & \texttt{CAP\_RECEIVE\_DATA} & Receive data from network connections & \texttt{recvfrom}, \texttt{recvmsg} \\
\midrule

\multirow{2}{*}{Execution Capabilities} 
 & \texttt{CAP\_EXEC} & Launch new processes or threads & \texttt{clone}, \texttt{execve} \\
 & \texttt{CAP\_TERMINATE\_PROCESS} & Terminate running processes or threads & \texttt{exit}, \texttt{kill} \\
\midrule

\multirow{3}{*}{\shortstack[l]{System State and\\Configuration}}
 & \texttt{CAP\_READ\_SYSTEM\_STATE} & Access system configuration or status information & \texttt{getpid}, \texttt{getitimer} \\
 & \texttt{CAP\_WRITE\_SYSTEM\_STATE} & Modify environment variables or system settings & \texttt{setuid}, \texttt{setgid} \\
 & \texttt{CAP\_RESOURCE\_LIMITS} & Adjust process or system resource limits & \texttt{setrlimit} \\
\midrule

\multirow{2}{*}{Memory Operations}
 & \texttt{CAP\_MEMORY\_MANIPULATION} & Alter memory regions or mappings at runtime & \texttt{munmap}, \texttt{mmap} \\
 & \texttt{CAP\_DIRECT\_IO} & Perform low-level I/O operations on devices or memory & \texttt{ioctl} \\
\bottomrule
\end{tabular}
\label{tab:capability_taxonomy}
\end{table*}

\subsection{Analysis Mode}

Once system calls are attributed to specific capabilities and the invoking package is identified through stack trace resolution, GoLeash aggregates this information to build a capabilities \emph{allowlist} for each package. These allowlists are constructed in the \emph{analysis mode}, where the application is executed in a trusted environment, using integration tests or representative workloads to exercise the intended behavior of the application. This generated allowlist acts as a behavioral contract: any future execution is expected to conform to these observed patters. Once reviewed and approved, the allowlist becomes the \emph{policy} used for enforcement.

% what is an allow list
For each traced package exercised during the execution, the allowlist records the set of capabilities it invoked, as well as all observed call paths that led to each capability. These call paths capture the broader execution context: they consist of the ordered sequence of Go packages extracted from the stack trace, with the traced package acting as the terminal caller responsible for the capability. 

\Def{
\label{def:allowlist}
Formally, the allowlist $\mathcal{A}$ is defined as:
\[
\mathcal{A} = \{ (P_i, \{ (C_{ij}, \mathcal{T}_{ij}) \}) \}
\]
where:
\begin{itemize}
    \item $P_i$ is a traced package,
    \item $C_{ij}$ is a capability invoked by $P_i$,
    \item $\mathcal{T}_{ij} = \{ T_{ijk} \}$ is the set of call paths observed for capability $C_{ij}$,
    \item each call path $T_{ijk}$ is an ordered list of Go packages:
    \[
    T_{ijk} = [Q_1, Q_2, \dots, Q_n, P_i]
    \]
    with $Q_k$ denoting intermediary packages and $P_i$ as the terminal traced package.
\end{itemize}
}

% hash, cool
To optimize for performance and memory efficiency, GoLeash stores each observed call path as a hash of the array $\mathcal{T}_{ij}$. This compact representation allows for fast lookups at run-time while preserving the ability to verify complete calling contexts. Recalling the example in Figure \ref{fig:stacktrace}, during analysis, GoLeash saves the entire sequence of legitimate call paths executed, such as \texttt{[cobra} $\rightarrow$ \texttt{golib} $\rightarrow$ \texttt{write]}.

 % process
\subsection{Policy Management in the Development Process}
\label{sec:policy_management}

% Why refinement is needed
GoLeash policies are refined over time. Two factors primarily motivate this iterative refinement. First, Go programs often rely on dynamic constructs, such as function pointers, interface-based dispatch, and asynchronous callbacks, that yield variability in execution. Some legitimate syscall-triggering paths may not be exercised during initial testing. Second, as software evolves, new features and updates can introduce new system behaviors not present during the original analysis. Major dependency upgrades, in particular, may bring new capabilities requiring policy updates.

% How refinement is achieved
To accommodate this, GoLeash policies are intentionally designed to be iterative and non-disruptive. The system builds a conservative baseline from observed behavior during analysis mode, which can then be incrementally extended later. When new legitimate paths are exercised, whether during testing or staged deployment, their associated capabilities can be appended to the existing policy without discarding prior entries.

Developers can also manually audit and adjust policies. For example, before production deployment, they may inspect which Go packages initiate sensitive actions, such as executing binaries or opening network connections, and freeze the allowlist after the review to prevent future unauthorized changes. In addition, certain capabilities are inherently security-sensitive and should be treated as explicit audit points during policy review. Capabilities enabling privileged process interaction or memory manipulation, for instance, are rarely required by typical application workloads and may substantially expand the attack surface if granted unnecessarily. Their presence in a generated policy therefore serves as a strong signal for manual inspection and justification prior to deployment.

In practice, developers typically update policies during major version upgrades of dependencies, where functional additions are more likely. Empirical studies on software and permission evolution demonstrate that the churn of system-level capabilities (e.g., system calls and permissions) across consecutive versions is generally low~\cite{calciati2017apps, bagherzadeh2018analyzing}. When new features are introduced (typically during major releases) the observable behavior of the code expands, occasionally requiring new capabilities. However, minor updates are usually limited to bug fixes and internal refactoring. Extensive empirical research on software evolution confirms that refactoring and bug-fixing strictly preserve existing APIs and observable system behaviors~\cite{dig2005role, bagherzadeh2018analyzing}, meaning the underlying capabilities remain unchanged. Consequently, policy churn is minimal across minor updates, and developers are not burdened with manual policy reviews unless a major version upgrade explicitly introduces new, privileged operations.

To sum up, GoLeash policies should be updated and re-audited whenever a dependency publishes a major version, or when new privileged operations are explicitly added in Changelogs.

\subsection{Enforcement Mode}
\label{subsec:enforcement}
Once a \emph{policy} has been constructed and approved during development, GoLeash can be used in the enforcement mode to protect the running application from unauthorized behaviors. The same eBPF-based tracing infrastructure is reused to intercept system calls; however, instead of gathering these observations for analysis, GoLeash now checks each syscall in real-time against the policy. Enforcement is made at runtime by validating both the origin of the syscall and the capability being exercised.

For every intercepted syscall, GoLeash resolves its associated capability and identifies the terminal package responsible, using the same attribution and classification logic from analysis. The syscall is then validated against the \emph{policy}: if the \texttt{(package, capability)} pair is present, and the corresponding call path matches one of the approved sequences stored for that capability, the syscall is allowed to proceed uninterrupted. Otherwise, GoLeash flags the event as a policy violation and triggers a configurable enforcement action.

By validating not only the invoking package but also the full call path in the stack trace, GoLeash supports context-aware enforcement. This defends against confused deputy attacks~\cite{cwe441}, when a restricted dependency may attempt to trigger a privileged capability indirectly by invoking a more permissive package within the same application. In such cases, even if the terminal package is authorized for a capability, the unapproved calling context causes the request to be denied. In the example of Figure \ref{fig:stacktrace}, an attacker could compromise \texttt{spf13/cobra} to import a function from \texttt{xtaci/kcp}, triggering a \texttt{SEND\_DATA}, which was not part of \texttt{cobra}'s original capability set. GoLeash prevents the invocation path between cobra and \texttt{kcp}, since it is not part of the policy.

GoLeash supports multiple enforcement strategies, allowing it to be used for forensics and response purposes. In forensics use cases, violations are logged for postmortem analysis, enabling developers to assess suspicious behaviors without disrupting application functionality. In response use cases, GoLeash can terminate the offending process entirely, preventing potential exploitation in real time. These strategies can be configured per environment, supporting progressive hardening.

\subsection{Advanced Support}
\paragraph{Support for Trusted Go Internals}  
It is important to consider that the Go runtime itself can initiate system calls, independently from application logic, such as, for scheduling, I/O polling, and garbage collection. 
GoLeash distinguishes between system calls originating from the Go runtime from those issued by packages of the application and modules. 
These runtime-initiated events are excluded from enforcement, by examining the stack trace associated with each system call. If the entire call stack contains only frames from Go’s runtime and does not include any package of the application or modules, the system call is treated as coming from trusted infrastructure and excluded from analysis. This design ensures that only behavior explicitly caused by the application or its dependencies is subject to capability enforcement, reducing the overhead and avoiding false positives.

\paragraph{Support for Evasion Techniques} 
The syscall attribution mechanism implemented in GoLeash is robust even in the presence of advanced attack vectors that use defense evasion techniques, including obfuscation. Techniques such as encoded Go code, CGO bindings, inline assembly, and dynamically loaded plugins may attempt to execute malicious operations. However, all such operations ultimately result in syscalls that pass through the kernel. Since the stack includes the originating Go package, GoLeash can identify the dependency involved in the syscall. Moreover, GoLeash is able to handle malicious code that hides its behavior by executing existing binaries on the target system (\emph{living-off-the-land} attacks), as in MITRE ATT\&CK \href{https://attack.mitre.org/techniques/T1036/}{\textit{T1036 (Masquerading)}}, as well as through obfuscation and control flow manipulation strategies commonly seen in \href{https://attack.mitre.org/techniques/T1574/}{\textit{T1574 (Hijack Execution Flow)}}, as described later.

\paragraph{Support for Exec-based Control Transfer} 
Go applications commonly invoke external binaries to delegate tasks, such as running system utilities, interacting with non-Go components, and launching helper tools. This pattern is supported directly by the Go standard library via wrappers around \texttt{exec()}. However, this behavior introduces challenges for security monitoring. The \texttt{exec()} syscall replaces the current process image with a new binary while retaining the same PID, effectively shifting execution to code that may not be vetted or trusted. Attackers can exploit this to bypass runtime controls from within a compromised package. This tactic aligns with MITRE ATT\&CK techniques \href{https://attack.mitre.org/techniques/T1055/}{\textit{T1055 (Process Injection)}} and \href{https://attack.mitre.org/techniques/T1543/}{\textit{T1543 (Create or Modify System Process)}}. To handle this safely, GoLeash analyzes both \texttt{sys\_enter} and \texttt{sys\_exit} events for the same \texttt{exec()} syscall. A pending ``external binary execution'' event is recorded at the syscall entry, and only committed if the syscall completes successfully. This avoids recording syscalls from failed or aborted transitions. It must be noted that the external binary is not necessarily a Go program. Thus, after a successful exec, GoLeash tracks syscalls using a flat (i.e., dependency-unaware) allowlist for the process as a whole, indexed by the new executable's name. During this transition, residual syscalls, such as those issued by the Go runtime or by parallel threads active at the moment of the \texttt{exec()}, are filtered out.

\paragraph{Support for Multi-process and Multi-Binaries} 
GoLeash handles runtime behaviors that complicate syscall attribution in Go, such as process forking and multi-binary applications. Supporting these behaviors is essential, as they are common in modern Go systems. Forking is used to spawn workers, isolate tasks, and handle concurrent requests, while multi-binary applications assign distinct responsibilities to separate executables. This design is especially prevalent in distributed systems. For example, Kubernetes runs several binaries within a single node, and often deploys multiple replicas of the same binary for redundancy and load balancing. These behaviors are also security-relevant. Attackers may fork subprocesses to evade runtime monitoring or enforce stealth, aligning with MITRE ATT\&CK techniques such as \href{https://attack.mitre.org/techniques/T1059/}{\textit{T1059 (Command and Scripting Interpreter)}}. GoLeash counters this by dynamically tracking all processes spawned across all binaries in the target application, including those created via \texttt{fork()}, ensuring that syscall attribution and policy enforcement remain consistent across the entire process tree.

% -------------------------------------------------------------------------------------------%
\subsection{Implementation Details}
We implemented GoLeash using the C and Go languages. The codebase includes $1,190$ lines of code, split between the eBPF tracing component (kernel-space) and the Go-based analysis and enforcement engine (user-space).

The kernel component uses the \texttt{cilium/bpf2go} library \cite{bpf2go} to attach probes that capture syscall events. Each event includes a hash-based ID derived from the user-space stack trace, enabling efficient, deduplicated lookup for execution paths. Events are relayed to user space via a ring buffer to ensure low-overhead, non-blocking communication. The user-space component resolves stack traces, maps instruction addresses to Go symbols and Go package paths, and enforces capability policies. This process leverages symbol information embedded in Go binaries, which includes fully qualified package paths, allowing for accurate attribution without source code access. 
GoLeash targets x86-64 Linux systems and requires a kernel version 5.7 or later for full eBPF tracing support.

% -------------------------------------------------------------------------------------------%

\section{Experimental Analysis}
\label{sec:evaluation}

We conduct a systematic, large-scale experimental evaluation of GoLeash, structured around the following research questions.

\begin{description}
    \item[RQ1:] \textbf{How effectively does GoLeash detect malicious code in third-party dependencies?}
    We investigate GoLeash's ability to identify unauthorized use of system capabilities as part of software supply chain attacks. The experimental protocol consists in simulating realistic supply chain attacks, by injecting malicious behaviors into benign Go projects, and measuring GoLeash's detection rate.
    
    \item[RQ2:] \textbf{How robust is GoLeash in detecting obfuscated malicious behavior?}
    Typically, malicious software is hidden through obfuscation techniques, to make it more difficult to detect by malware detectors. Thus, we assess whether obfuscation techniques can evade GoLeash's enforcement mechanism.

    \item[RQ3:] \textbf{What is the performance overhead introduced by GoLeash?}
    We measure the overhead introduced by GoLeash in terms of system call latency and overall execution time of traced applications. 
\bigbreak
    \item[RQ4:] \textbf{How does GoLeash compare with static capability analysis?}
    GoLeash leverages dynamic analysis for capability attribution, complementing static approaches. Therefore, we quantify the overlap and differences between GoLeash and a state-of-the-art static capability analysis tool.

    %\item[RQ5:] \textbf{Can GoLeash be used to mitigate runtime exploitation beyon SSC attacks?}
    %We explore GoLeash's effectiveness in blocking runtime exploits such as remote code execution (RCE).
\end{description}

\bigbreak
\textit{Experimental Targets.} 
We evaluate GoLeash on real-world, complex Go software projects. We started from the GitHub Search API, which lists ~2.3M public repositories with Go as their primary language as of May 2025. We first applied two automatic filters: (1) more than 10{,}000 stars, and (2) active maintenance, defined as at least one commit in the last six months. This produces a shortlist of 371 candidate projects. From this shortlist, we manually selected projects that (3) ship a test suite or workload exercising their core functionalities, and (4) are security sensitive, i.e., projects whose compromise would have direct security implications for dependent systems. Applying this strict criterion results in the following five projects:
\textit{kubernetes} (k8s) \cite{kubernetes} for container orchestration, 
\textit{etcd}~\cite{etcd} for key-value storage, \textit{coredns}~\cite{coredns} and \textit{frp} (fast reverse proxy)~\cite{frp} for networking, and \textit{go-ethereum} (geth)~\cite{goethereum} for blockchain infrastructure. 

Those applications are complex and some contain multiple binaries.
We include all binaries in our evaluation. 
For example, in the case of Kubernetes, we analyzed the five main control plane binaries: \textit{kube-apiserver}, \textit{kube-controller-manager}, \textit{kube-scheduler}, \textit{kube-proxy}, \textit{kubelet}.

\bigbreak
\textit{Malicious Code Injection.} 
Malicious behaviors in supply chain attacks often abuse system-level capabilities such as establishing outbound connections to exfiltrate data, accessing files to steal sensitive content, or executing system commands to escalate access or evade detection.

To the best of our knowledge, no publicly available dataset exists for malicious Go packages. While such packages do appear in the wild and represent an active threat, it is a known problem \cite{williams2025research} that they are usually removed from public repositories shortly after detection, impeding reproducible research.
We tried to access the few malicious ones reported in blog posts, but did not succeed in obtaining the code.
For our evaluation, we tackle this lack of data by building synthetic malicious packages according to the following sound methodology.

First, we surveyed two established datasets of long-lived supply chain attacks across npm, PyPI, RubyGems, and Packagist~\cite{ohm2020backstabber,datadog_malicious_dataset}. We grouped the documented payloads by capability profile, that is, which OS-level capabilities they exercise: filesystem access, network egress, process execution, and native code loading.

Second, from each capability cluster we selected one widely observed behavior pattern, yielding five malware families.

Third, we ported each behavior to Go using the language's standard library and APIs commonly abused in real Go-ecosystem attacks~\cite{go_supply_chain_attack_boltdb,hypert,go_evil_packages}.

Each pattern preserves the system-level interaction pattern of the original incident while substituting Go-native APIs for those of the source ecosystem (e.g., \texttt{os/exec.Command} in place of Node's \texttt{child\_process.exec}).
The injected payloads are therefore not arbitrary author constructions: each one mirrors a documented attack pattern at the granularity at which GoLeash operates.
Table~\ref{tab:malicious_behavior} maps each behavior to its real-world inspiration, and to the relevant entries of the MITRE ATT\&CK knowledge base.

Then, for each evaluated Go application, we first profiled its benign execution under standard workloads to identify which of its imported packages are actually exercised at runtime. This ensures malware injection targets only reachable, active parts of the application’s dependency graph.

We then injected each of the five malware behaviors into \emph{every exercised package}, one package at a time. This exhaustively explores the runtime attack surface. 

We automate the injection of malicious code by rewriting the source code of the target package. We developed a static rewriting tool that analyzes the abstract syntax tree of the package and weaves a malicious snippet into an existing function without altering the original control flow or breaking functionality. For each package, the injection point is randomly selected from the set of functions confirmed to execute during profiling. This ensures realism while maintaining reproducibility and isolation of injected behavior.

\begin{table}[t]
\centering
\caption{Injected malicious behaviors.}
\label{tab:malicious_behavior}
\rowcolors{2}{gray!10}{white}
\begin{tabular}{
    l
    >{\raggedright\arraybackslash}p{0.3\linewidth}
    >{\raggedright\arraybackslash}p{0.10\linewidth}
    >{\raggedright\arraybackslash}p{0.28\linewidth}
}
\toprule
\textbf{ID} & \textbf{Behavior} & \textbf{MITRE ATT\&CK} & \textbf{Inspired by} \\
\midrule

M1 & Base64-encoded command execution for reverse shell access &
\href{https://attack.mitre.org/techniques/T1059/}{T1059},
\href{https://attack.mitre.org/techniques/T1140/}{T1140} &
48 npm reverse-shell packages~\cite{phylum_npm_reverse_2023} \\

M2 & Exfiltrate and transmit system configuration data to a remote server &
\href{https://attack.mitre.org/techniques/T1552/}{T1552},
\href{https://attack.mitre.org/techniques/T1082/}{T1082},
\href{https://attack.mitre.org/techniques/T1041/}{T1041} &
PyTorch-nightly \texttt{torchtriton}~\cite{pytorch_torchtriton_2022} \\

M3 & Information stealing from user applications &
\href{https://attack.mitre.org/techniques/T1555/}{T1555},
\href{https://attack.mitre.org/techniques/T1005/}{T1005} &
\texttt{ua-parser-js} npm compromise~\cite{mandiant_uaparser_2021} \\

M4 & Inject a shared library into current process &
\href{https://attack.mitre.org/techniques/T1129/}{T1129} &
\texttt{termncolor} and \texttt{colorinal} PyPI~\cite{zscaler_termncolor_2025} \\

M5 & Infiltrate malicious files &
\href{https://attack.mitre.org/techniques/T1105/}{T1105} &
\texttt{klow}, \texttt{klown}, and \texttt{okhsa} npm cryptominers~\cite{sonatype_klow_2021} \\
\bottomrule
\end{tabular}
\end{table}

\bigbreak
\textit{Policy Scalability.}
In analysis mode, GoLeash generates one policy per application, structured as a set of entries, each entry describing the capabilities exercised by a single Go package. Across our five benchmarks, the number of entries per policy ranges from 20 to 441 (e.g., 441 exercised packages in Kubernetes). On disk, these policies remain compact, with an average size of approximately 90\,KB. Capability usage within each policy is sparse: for instance, in Kubernetes the median number of capabilities per package is 3, which makes the resulting policies practical to manually inspect and audit. The syscall-to-capability mapping is maintained as a single JSON file covering 456 Linux syscalls. 
Since the Linux syscall interface mainly evolves by adding new calls, supporting newer kernel versions only requires appending the newly introduced syscalls to the existing JSON mapping and assigning them to capabilities. This is a one-time, linear update per kernel version.

\bigbreak
\textit{System Configuration.} 
All experiments are conducted on a machine with an Intel Core i7-1260P CPU, 16GB of RAM, running Ubuntu 22.04, with Linux kernel version 6.7, and Go version 1.24.1.

\subsection{RQ1: Effectiveness Against Malicious Behavior in Dependencies}
\label{subsec:rq1}

\textit{Objective.}
The goal of this experiment is to evaluate GoLeash’s effectiveness in mitigating supply chain attacks by detecting unauthorized system capability usage introduced through compromised third-party packages. 
We aim to show that GoLeash's per-package enforcement can accurately flag malicious behaviors, even when injected into deep transitive dependencies. 
To contextualize GoLeash’s effectiveness, we compare it against a application-level enforcement baseline, which represents the granularity enforced by common Mandatory Access Control (MAC) mechanisms like SELinux \cite{selinux} and AppArmor \cite{apparmor}, as well as syscall filtering tools like Seccomp \cite{seccomp_linux}.
A detection is considered successful when GoLeash successfully detects that a malicious package exercises a malicious capability absent from the policy.

\bigbreak
\textit{Methodology.}
% Policy Generation
We first use GoLeash in \emph{analysis mode} to generate per-package policies. This is done by executing each target application with its end-to-end workload while GoLeash traces all invoked packages. These workloads are designed to reflect the application's intended, benign behavior. 

% Policy generation for Baseline 
To compare GoLeash’s per-package enforcement with traditional sandboxing models, we implement a \emph{\textbf{application}-level enforcement baseline}. This baseline aggregates all observed capabilities into a flat, application-wide policy, without attributing them to individual packages. This setup mirrors the enforcement granularity of MAC systems like SELinux and AppArmor, which operate at the \textbf{application} level. Since no ready-to-use profiles exist for our target applications, and manual policy creation is error-prone and labor-intensive, our automatically generated application-level policy serves as a practical, reproducible approximation of what such systems can achieve in practice.

% Malicious Application Dataset 
After policy generation, we construct a dataset of malicious variants of the target applications. For each variant, we rebuild the application with exactly one malicious code snippet at a time, injected into a single target package. The injection point is randomly selected from functions that are actually executed, which are profiled using our tracing infrastructure. It is worth noting that not all packages in the dependency graph are actually reachable by the workload. For example, the \texttt{net} Go package may import \texttt{crypto} for encrypted communication, but the application using \texttt{net} might not use encryption features. Therefore, we limit injections to packages that are actually active during execution. This ensures that the malicious code is reachable and that it represents a realistic attack. We leverage the injection approach described in Section~\ref{sec:evaluation}. In total, we produced \emph{3,265} unique malicious application variants.

% Metric for comparison
The effectiveness of GoLeash is measured using the \emph{detection rate}, defined as the percentage of malicious packages flagged by GoLeash out of the total number of created malicious packages for a given application. For comparison, the same metric is applied to the application-level baseline. Under this coarse-grained model, a variant is only detected if the injected code uses capabilities not observed anywhere in the application during baseline execution.

Malware IDs used in the evaluation are defined in Table~\ref{tab:malicious_behavior}, and the \textit{inj.} column indicates the number of application packages individually injected with each malware for testing.

% GoLeash is able to detect 98\% of software supply chain attacks.
\begin{table}
\centering
\caption{Detection Rate of GoLeash vs App-level. Malware IDs are defined in Table~\ref{tab:malicious_behavior}. The \textit{``inj.''} column indicates the number malicious application variants.}
\begin{tabular}{llccc}
\toprule
\textbf{Project} & \textbf{Malw.} & \textbf{Inj. (\#)} & \textbf{App-level (\%)} & \textbf{GoLeash (\%)} \\
\midrule
\multirow{5}{*}{\texttt{k8s} \cite{kubernetes}}
 & M1 & 441 & 100 & 100 \\
 & M2 & 441 & 0   & 99.77 \\
 & M3 & 441 & 100 & 100 \\
 & M4 & 441 & 0   & 99.32 \\
 & M5 & 441 & 0   & 99.55 \\
\midrule
\multirow{5}{*}{\texttt{etcd} \cite{etcd}}
 & M1 & 62  & 100 & 100 \\
 & M2 & 62  & 0   & 98.39 \\
 & M3 & 62  & 0   & 100 \\
 & M4 & 62  & 0   & 91.94 \\
 & M5 & 62  & 0   & 98.39 \\
\midrule
\multirow{5}{*}{\texttt{coredns} \cite{coredns}}
 & M1 & 20  & 100 & 100 \\
 & M2 & 20  & 0   & 100 \\
 & M3 & 20  & 100 & 100 \\
 & M4 & 20  & 0   & 95 \\
 & M5 & 20  & 0   & 100 \\
\midrule
\multirow{5}{*}{\texttt{frp} \cite{frp}}
 & M1 & 47  & 100 & 100 \\
 & M2 & 47  & 0   & 97.87 \\
 & M3 & 47  & 100 & 100 \\
 & M4 & 47  & 0   & 87.23 \\
 & M5 & 47  & 0   & 95.74 \\
\midrule
\multirow{5}{*}{\texttt{geth} \cite{goethereum}}
 & M1 & 83  & 100 & 100 \\
 & M2 & 83  & 0   & 97.59 \\
 & M3 & 83  & 0   & 100 \\
 & M4 & 83  & 0   & 91.57 \\
 & M5 & 83  & 0   & 97.59 \\
\midrule
\textit{Aggregate} &  & 3,265 & 32\%  & 98\%  \\
\bottomrule
\end{tabular}
\label{tab:malware_detection}
\end{table}

\bigbreak
\textit{Results.}
Table~\ref{tab:malware_detection} summarizes the results of this experiment. 
As previously discussed, we profile which packages are actually executed (e.g., $441$ in the case of k8s), and inject five types of malicious code into each target package. In total, we obtain $3,265$ malware injections.

GoLeash successfully detects malicious behavior with high accuracy. The average detection rate across all projects is $98\%$. This is much higher than the $32\%$ detection rate achieved by the application-wide baseline approach. GoLeash achieves a perfect $100\%$ detection rate on two different malware categories: M1 and M3. Importantly, the detection rate never dropped below $87\%$.

GoLeash effectively identifies deviations from normal execution, by detecting when the injected malicious code uses capabilities that are not part of the policy generated with benign executions. Our results indicate that certain sensitive capabilities, such as initiating outbound network connections or spawning external commands, used by malware, are particularly effective indicators.

In contrast, the coarse-grained, application-level baseline model was significantly less effective. By aggregating all capabilities into a single policy, application-level enforcement often allows stealthy malware to go undetected. This happens when the malicious code uses capabilities that are also legitimately used by the target package. 
The baseline is able to detect malware M1 and M3, which leverages the rarely-used capability \texttt{RESOURCE\_LIMITS} capability, which is absent in the original applications.

\Definition{We have performed a large-scale experiment with 3265 malicious packages injected in five real-world Go projects. GoLeash stops 98\% of the simulated software supply chain attacks. GoLeash’s per-package enforcement offers precise and robust defense against software supply chain attacks. 
}

\subsection{RQ2: Effectiveness Against Obfuscated Attacks}
\label{subsec:rq2}

\textit{Objective.}
Malicious actors often attempt to hide dangerous logic via obfuscation techniques, making it harder for security tools to detect unauthorized actions \cite{xu2012power, o2011obfuscation}. This experiment evaluates whether obfuscation techniques impact the ability of GoLeash to perform per-package capability enforcement. 

\bigbreak
\textit{Methodology.} 
As in RQ1, we begin by running each target application with its normal workload in GoLeash analysis mode, generating a per-package policy that captures legitimate capabilities.
Then, we inject obfuscated malicious code,  using four Go-specific obfuscation strategies \cite{cesarano2023gosurf}:  

\begin{itemize}[leftmargin=1.2em, itemsep=0pt]
    \item \textbf{Plugin}: we compile the malicious functionality as a Go plugin, and load it at runtime.
    \item \textbf{Reflection}: we wrap malicious functions with reflection calls, with dynamic resolution to invoke them.
    \item \textbf{External Binary}: we embed the payload in a separate binary, and invoke it at runtime via \texttt{exec} syscall.
    \item \textbf{CGO}: we write the malicious code as C code, and invoke it through Go’s foreign function interface (CGO).
\end{itemize}

Each of these obfuscation techniques conceals the malicious injected capabilities behind more convoluted, indirect execution paths. 

We reuse the same injection framework described in Section \ref{sec:evaluation}, but apply the above four evasion strategies to the malicious snippets before injecting them into each exercised package. In this second experiment, we focus on two malicious behaviors (i.e., M2 data exfiltration and M5 remote file infiltration) from the original set, for the sake of brevity. Similar results were obtained for the other types of malware. Each obfuscation variant is injected one at a time into each exercised package across the five Go projects. We then switch GoLeash to enforcement mode and rerun the applications with their workload to measure if GoLeash detects these obfuscated attacks.

\begin{table}[ht]
\centering
\caption{Detection Rate of GoLeash against obfuscated attacks}
\label{tab:obfuscation_detection}

\subfloat[{\footnotesize Malware M2 (data exfiltration)}\label{tab:obfuscation_detection_M2}]{
\begin{tabular}{lccccc}
\toprule
\textbf{Project} & \textbf{w/o(\%)} & \textbf{plug.(\%)} & \textbf{refl.(\%)} & \textbf{bin.(\%)} & \textbf{cgo(\%)} \\
\midrule
\texttt{k8s}     & 99.77 & 99.77 & 99.77 & 100 & 99.09 \\
\texttt{etcd}    & 98.39 & 98.39 & 98.39 & 100 & 98.39 \\
\texttt{coredns} & 100   & 100   & 100   & 100 & 90    \\
\texttt{frp}     & 97.87 & 93.62 & 93.62 & 100 & 76.60 \\
\texttt{geth}    & 97.59 & 97.59 & 97.59 & 100 & 96.39 \\
\midrule
\texttt{avg.}    & 98.72 & 97.87 & 97.87 & 100.00 & 92.09 \\
\bottomrule
\end{tabular}
}
\hfill
\subfloat[{\footnotesize Malware M5 (remote file infiltration)}\label{tab:obfuscation_detection_M5}]{
\begin{tabular}{lccccc}
\toprule
\textbf{Project} & \textbf{w/o(\%)} & \textbf{plug.(\%)} & \textbf{refl.(\%)} & \textbf{bin.(\%)} & \textbf{cgo(\%)} \\
\midrule
\texttt{k8s}     & 99.55 & 99.55 & 99.55 & 100 & 99.32  \\
\texttt{etcd}    & 98.39 & 98.39 & 98.39 & 100 & 98.39  \\
\texttt{coredns} & 100   & 100   & 100   & 100 & 95     \\
\texttt{frp}     & 95.74 & 95.74 & 95.74 & 100 & 85.11 \\
\texttt{geth}    & 97.59 & 97.59 & 100   & 96.39 & 96.39 \\
\midrule
\texttt{avg.}    & 98.25 & 98.25 & 98.33 & 99.28 & 94.44 \\
\bottomrule
\end{tabular}
}

\end{table}

\bigbreak
\textit{Results.}
Tables~\ref{tab:obfuscation_detection_M2} and \ref{tab:obfuscation_detection_M5} report GoLeash's detection rates of GoLeash for both unobfuscated malware as in RQ1 (column \texttt{w/o}), and their obfuscated counterparts. For each exercised package in the target applications, we generate eight malicious injections (two malware variants and four evasion techniques), resulting in a total of $5,224$ injections across all five applications.

Despite attempts to conceal malicious capabilities, GoLeash still identifies unauthorized behavior that falls outside a package’s generated policy, achieving average detection rates across all projects ranging from $92.09\%$ to $100\%$ for malware M2, and from $94.44\%$ to $99.28\%$ for malware M5, depending on the obfuscation technique. Our experiment shows that the evasion strategies poorly circumvent GoLeash’s per-package policies.
Malicious accesses are still flagged as suspicious calls whenever they deviate from a package’s baseline profile. 
For example, malware based on CGO still needs to issue system calls for connecting to remote hosts, such as those mapped to the \texttt{\texttt{CONNECT\_REMOTE}} capability, which remains detectable. 
Interestingly, in the scenario of external binary evasion, GoLeash’s detection rate is slightly higher than in the unobfuscated variant, primarily because this approach introduces the \texttt{EXEC} capability, which is absent in most of the baseline profile and immediately triggers an alert.

\Definition{
GoLeash is highly robust against obfuscated software supply chain attacks, where the malicious code in obfuscated inside a dependency. Despite obfuscation, GoLeash detects the deviations from the security policy.}

\subsection{RQ3: Performance Overhead}

\textit{Objective.} 
The goal of this evaluation is to quantify the runtime overhead introduced by GoLeash when used in enforcement mode. Recall that the enforcement mode is meant to be used in production. In enforcement mode, GoLeash actively intercepts and potentially blocks system calls from the target application. Since runtime overhead can impact real-world adoption, it is critical to measure how much additional cost is imposed by GoLeash.

\bigbreak
\textit{Methodology.}
Each evaluated application runs as a long-running server process driven by an external client that generates a representative request-driven workload; we measure overhead on the server side. The five workloads, listed below, replace the functional end-to-end workloads used in RQ1.

For \texttt{etcd}, we run its official benchmarking tool~\cite{etcd_benchmark} with $200{,}000$ \texttt{put} requests against a single-node \texttt{etcd} server. 
For \texttt{CoreDNS}, we issue DNS queries using \texttt{DNSPerf}~\cite{dnsperf} over a $30$-second sustained window.
For \texttt{frp}, we tunnel HTTP traffic through the frp proxy to a backend web server and drive it with \texttt{wrk}~\cite{wrk} for $30$ seconds.  
For \texttt{geth}, we drive a local Ethereum node with concurrent read requests for $30$ seconds.
For \texttt{Kubernetes}, we drive \texttt{kube-apiserver} with concurrent API read requests (GET \texttt{/api/v1/pods}) for $30$ seconds.

For each application we run the benchmark $20$ times in three conditions. First, we run the application without GoLeash, as a baseline. Second, we measure performance with GoLeash attached to the server in enforcement mode. Third, to isolate the contribution of stack-trace resolution to the cost, we additionally measure the \emph{application-level} variant of enforcement without walking the user stack on every syscall. The target application, the load generator, and the eBPF userspace consumer are pinned to disjoint CPU sets, and the page cache is dropped between iterations. We report median values after $1.5{\times}$IQR outlier removal, and assess statistical significance using Welch's t-test.

We report two complementary metrics per condition. 
The \emph{CPU overhead per request} (\%) is the relative increase in server's average CPU time per processed request, computed against the GoLeash-free baseline; it is workload-normalized, independent of the application's CPU utilization regime. 
The added \emph{syscall latency} (ns/syscall) is the additional CPU time the probe consumes per traced syscall, measured as the absolute additional CPU time per request (in ns) divided by the number of system calls the workload issues per request; it isolates the probe's intrinsic kernel-side cost from each workload's syscall density and is directly comparable across applications.

\begin{table}[ht]
\centering
\setlength{\tabcolsep}{4pt}
\caption{Performance overhead of GoLeash and its application-level variant, relative to the GoLeash-free baseline.}
\label{tab:overhead}
\begin{tabular}{lcccc}
\toprule
& \multicolumn{2}{c}{\textbf{CPU Overhead Per-request (\%)}}
& \multicolumn{2}{c}{\textbf{Syscall Latency (ns)}} \\
\arrayrulecolor{black!40}
\cmidrule(lr){2-3}
\cmidrule(lr){4-5}
\arrayrulecolor{black}
\textbf{Project}
& \textbf{GoLeash}
& \textbf{App-level}
& \textbf{GoLeash}
& \textbf{App-level} \\
\midrule
\texttt{etcd}       & $1.3^{***}$  & $0.5^{*}$    & $185^{***}$ & $70^{*}$ \\
\texttt{CoreDNS}    & $2.1^{***}$  & $0.3^{*}$    & $275^{***}$ & $42^{*}$ \\
\texttt{frp}        & $5.1^{***}$  & $1.6^{***}$  & $306^{***}$ & $95^{***}$ \\
\texttt{geth}       & $1.3^{***}$  & $0.2^{*}$    & $354^{***}$ & $64^{*}$ \\

\texttt{Kubernetes} & $2.1^{***}$ & $0.9^{***}$ & $1688^{***}$ & $775^{***}$ \\
\hline
\textbf{Range}      & $1.3$--$5.1$ & $0.2$--$1.6$ & $185$--$354^{\dagger}$ & $42$--$95^{\dagger}$ \\
\hline
\multicolumn{5}{l}{\scriptsize $^{\dagger}$\,ns range covers the four request servers; Kubernetes is a CPU-bound outlier (see text).} \\
\multicolumn{5}{l}{\scriptsize $^{***}\,p<0.001$, $^{*}\,p<0.05$.} \\
\end{tabular}
\end{table}

\bigbreak
\textit{Results.} 
Table~\ref{tab:overhead} summarizes the overhead measurements. In enforcement mode, GoLeash adds between $1.3\%$ and $5.1\%$ of CPU per request across the five applications, with all deltas statistically significant ($p<0.001$). For the four request-serving applications (\texttt{etcd}, \texttt{CoreDNS}, \texttt{frp}, \texttt{geth}), normalizing by syscall density yields a tight added syscall latency of $185$--$354$\,ns, dominated by the probe's fixed kernel-side cost. \texttt{Kubernetes} is the exception at $1688\,$ns/syscall ($775\,$ns in app-level mode): its \texttt{kube-apiserver} performs $\approx 1.8$\,ms of CPU per request, an order of magnitude more than the other services, the bulk of it in response serialization rather than syscalls. Thus, the same modest $2.1\%$ relative overhead corresponds to a much larger per-syscall figure. This indicates that the per-syscall cost combines a probe-intrinsic component with an application-dependent one that grows with per-request CPU intensity. In relative terms, however, the cost stays bounded: $\le 2.1\%$ CPU per request ($\le 0.9\%$ app-level) even for the heaviest service evaluated. The application-level variant lowers the cost to $0.2$--$1.6\%$ CPU per request, roughly three to four times cheaper than full enforcement on average. These results are consistent with previously published characterizations of eBPF-based tracing systems~\cite{craun2024eliminating}.

Across all five projects, no false-positive violations were observed during enforcement under the benign workloads of RQ3. This is by construction (analysis and enforcement use the same workload). Generalization to disjoint workloads is bounded by the workload-coverage limitation of dynamic analysis, discussed with concrete mitigations in Section~\ref{sec:discussion} (Workload Coverage).

\Definition{
GoLeash's enforcement-mode probe adds $1.3$--$5.1\%$ of server CPU per request across five production Go services. The lighter application-level variant reduces this to $0.2$--$1.6\%$ CPU. This is an acceptable cost in most production settings for security benefits provided against software supply chain attacks.}

\subsection{RQ4: Comparison with Static Analysis}
\textit{Objective.}
Our goal in this experiment is to assess how GoLeash compares to \textit{Capslock} \cite{capslock}. Capslock is a state-of-the-art static capability analysis tool for Go, developed by Google. Capslock obtains the transitive dependency graph of the application under analysis. Then, for each package used by the application, it infers whether the package can manipulate files, open network connections, or perform certain system-level operations. It scans the source code by means of static analysis, by looking for calls to the Go standard library in the call graph. These capabilities are similar to those used by GoLeash (see Table \ref{tab:capability_taxonomy}), even if at a coarser level of granularity. We investigate whether Capslock's static analysis can effectively detect unobfuscated and obfuscated malicious code, how it compares to GoLeash's runtime-based detection, and whether the two techniques succeed and fail under the same circumstances, so as to clarify whether they are competing or complementary.

\bigbreak
\textit{Methodology.}
We compare the two tools along two distinct axes, \emph{visibility} and \emph{detection}, without conflating them.

\begin{figure}[t]
    \centering
    \includegraphics[width=\linewidth]{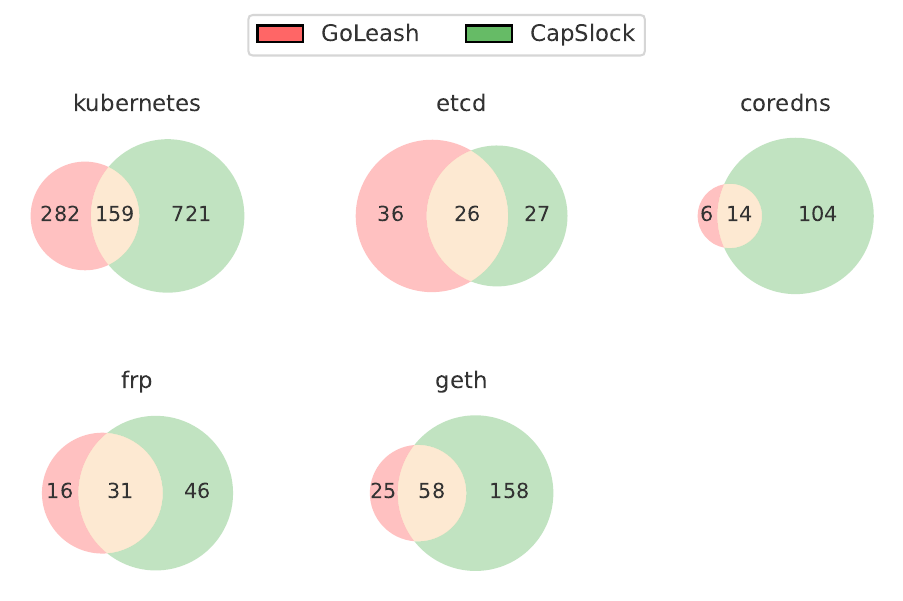}
    \caption{Package \emph{visibility}: number of packages uniquely reachable by GoLeash (dynamic) and Capslock (static), and their overlap, across five Go applications.}
    \label{fig:goleash_capslock_venn}
\end{figure}

\emph{Visibility} is the set of packages a tool can analyze at all. Capslock builds the transitive dependency graph and reasons about every imported package, including those never invoked at runtime, which increases coverage but incurs over-approximation. GoLeash, in contrast, observes only the packages actually exercised by the workload, yielding precise coverage of real execution but omitting unused code paths. The two visible sets therefore differ; Figure~\ref{fig:goleash_capslock_venn} quantifies their size and overlap for five analyzed applications. This axis concerns reachability only and says nothing about whether an injection is caught.

\emph{Detection} is whether, given that a package is visible to a tool, that tool flags an injection in it. We reuse the injected malicious packages from RQ1 and RQ2. A package is said to \emph{detect} an injection if the capabilities required by the malicious code are not already a subset of those the tool attributes to that package, i.e., the injection introduces at least one capability that the host package does not normally exhibit.

Because the two visible sets differ, a raw rate comparison would not be consistently fair to either tool; we therefore measure detection at two levels. \emph{(i)} A matched-coverage \emph{detection rate} computed on the \emph{exercised} packages identified by GoLeash, restricting Capslock to that same package set; the resulting differences are reported in Table~\ref{tab:detection_difference}. This is a deliberately conservative choice for GoLeash, since it discards Capslock's coverage of unexercised code. \emph{(ii)} A per-package \emph{agreement} analysis over the full population of Figure~\ref{fig:goleash_capslock_venn}, reported in Table~\ref{tab:rq4_agreement}, which decomposes outcomes by visibility region and additionally credits Capslock for the imported-but-unexercised packages only it can reach.

\begin{table}[ht]
\centering
\caption{Matched-coverage detection-rate difference (GoLeash $-$ Capslock) on the exercised packages.}
\label{tab:detection_difference}

\subfloat[{\footnotesize Malware M2 (Data Exfiltration)}\label{tab:detection_difference_m2}]{
\begin{tabular}{lccccc}
\toprule
 \textbf{Project} & \textbf{w/o(\%)} & \textbf{plug.(\%)} & \textbf{refl.(\%)} & \textbf{bin.(\%)} & \textbf{cgo(\%)} \\
\midrule
\texttt{k8s}     & +76.64 & +67.12 & +80.95 & +64.85 & +63.04 \\
\texttt{etcd}    & +62.91 & +58.07 & +69.36 & +58.06 & +56.45 \\
\texttt{coredns} & +50.00 & +50.00 & +60.00 & +45.00 & +20.00 \\
\texttt{frp}     & +70.21 & +29.79 & +70.22 & +36.17 & +10.64 \\
\texttt{geth}    & +62.65 & +32.53 & +67.47 & +33.73 & +27.72 \\
\midrule
\texttt{avg.} & +64.88 & +47.90 & +69.20 & +47.96 & +35.97 \\
\bottomrule
\end{tabular}
}
\hfill
\subfloat[{\footnotesize Malware M5 (Remote File Infiltration)}\label{tab:detection_difference_m5}]{
\begin{tabular}{lccccc}
\toprule
 \textbf{Project} & \textbf{w/o(\%)} & \textbf{plug.(\%)} & \textbf{refl.(\%)} & \textbf{bin.(\%)} & \textbf{cgo(\%)} \\
\midrule
\texttt{k8s}     & +74.38 & +66.90 & +80.73 & +64.85 & +63.27 \\
\texttt{etcd}    & +59.68 & +58.07 & +69.36 & +58.06 & +56.45 \\
\texttt{coredns} & +45.00 & +50.00 & +60.00 & +45.00 & +25.00 \\
\texttt{frp}     & +53.19 & +31.91 & +72.34 & +36.17 & +19.15 \\
\texttt{geth}    & +56.63 & +32.53 & +67.47 & +33.73 & +27.72 \\
\midrule
\texttt{avg.} & +57.38 & +47.88 & +69.58 & +47.96 & +38.72 \\
\bottomrule
\end{tabular}
}

\end{table}

\bigbreak
\textit{Results.}  
For space limitations, we again focus on two malware types. Tables~\ref{tab:detection_difference}a and \ref{tab:detection_difference}b report the absolute differences in detection rates between GoLeash and Capslock, respectively for the M2 and M5 malware types, across the five Go projects, on the exercised packages on which both tools are placed on equal footing. We consider both the cases without (\texttt{w/o}) and with obfuscation. Each cell shows the percentage-point difference between GoLeash's dynamic detection results from RQ2 and Capslock's static analysis on the same exercised packages.

GoLeash consistently and significantly outperforms Capslock across all projects and obfuscation techniques. On exercised packages its detection rate exceeds Capslock's by 64.9 points on average for unobfuscated M2 (Table~\ref{tab:detection_difference}a), and the advantage persists under every obfuscation technique, ranging from 36 to 69 average points. Results for M5 (Table~\ref{tab:detection_difference}b) are comparable. This gap stems from two main limitations of Capslock's analysis. First, Capslock operates at a coarser level of granularity than GoLeash, which causes overlap between broad capabilities attributed to trusted packages and those required by injected malware. As a result, it often fails to distinguish malicious behavior from benign behavior. For example, read and write operations fall in the same capability in Capslock, obscuring distinctions that GoLeash can capture at the syscall level. Second, Capslock's static analysis struggles to precisely reconstruct call graphs, especially in the presence of dynamic features such as plugin loading or reflection. Based on manual inspection of randomly sampled cases, we observed that Capslock occasionally misses even coarse-grained capabilities that are clearly exercised during runtime.

\bigbreak
\textit{Disagreement analysis.} To understand whether the two approaches fail under the same circumstances, we classify every package by visibility region and detection outcome in Table~\ref{tab:rq4_agreement} (shown for M2, \texttt{w/o} for the sake of brevity). Whereas Table~\ref{tab:detection_difference} reports an aggregate rate and Figure~\ref{fig:goleash_capslock_venn} reports reachability, the objective of Table~\ref{tab:rq4_agreement} is to expose, region by region, which tool catches which injection. Under the \emph{Analyzed by both} heading, the four columns sum to the overlap of Figure~\ref{fig:goleash_capslock_venn}; the \emph{Analyzed by one tool} columns report detected/visible in each tool's exclusive region.

On jointly-visible code, GoLeash strictly dominates. Within the shared region of Fig.~\ref{fig:goleash_capslock_venn}, GoLeash detects every injection Capslock detects plus a substantial set Capslock misses because of its coarser granularity (the \emph{GoLeash} column in Table~\ref{tab:rq4_agreement}). 
Conversely, across all five projects Capslock detects only a single shared package GoLeash misses (the \texttt{etcd} case in the \emph{Capslock} column), and at most two shared packages are missed by both (\texttt{geth}). This precision advantage on commonly observed code is what drives the rate gap in Table~\ref{tab:detection_difference}.

The two tools also analyze different package populations, and then their strengths are complementary. GoLeash detects injections in packages Capslock cannot place in its static call graph, typically those reached through reflection or plugin loading, and does so almost exhaustively (Table~\ref{tab:rq4_agreement}, \emph{GoLeash} column under \emph{Analyzed by one tool}). Conversely, Capslock detects injections in packages that are imported but never exercised by the workload, and that GoLeash therefore never observes (\emph{Capslock} column). These are not false negatives of GoLeash but lie on code paths the workload does not trigger, hence out of scope for any runtime technique by construction; they are precisely where static analysis retains unique value, bounding the unexercised long tail of dependencies.

\begin{table}[t]
\centering
\caption{Per-package detection outcome (M2, \texttt{w/o} obfuscation), partitioned by visibility region and detecting tool.}
\label{tab:rq4_agreement}
\setlength{\tabcolsep}{4pt}
\resizebox{\columnwidth}{!}{%
\begin{tabular}{lrrrrcc}
\toprule
 & \multicolumn{4}{c}{\textbf{Analyzed by both}} & \multicolumn{2}{c}{\textbf{Analyzed by one tool}} \\
\cmidrule(lr){2-5}\cmidrule(lr){6-7}
\textbf{Project} & \textbf{Both} & \textbf{GoLeash} & \textbf{Capslock} & \textbf{None} & \textbf{GoLeash} & \textbf{Capslock} \\
\midrule
\texttt{k8s}     & 102/159 & 57/159 & 0/159 & 0/159 & 281/282 & 469/721 \\
\texttt{etcd}    &  21/26 &  4/26 & 1/26 & 0/26 &   36/36 &  22/27  \\
\texttt{coredns} &  10/14 &  4/14 & 0/14 & 0/14 &     6/6 &  88/104 \\
\texttt{frp}     &  13/31 & 17/31 & 0/31 & 1/31 &   16/16 &  27/46  \\
\texttt{geth}    &  29/58 & 27/58 & 0/58 & 2/58 &   25/25 & 119/158 \\
\bottomrule
\end{tabular}%
}
\end{table}

Interestingly, when CGO-based obfuscation is used, GoLeash still outperforms Capslock in every scenario, although the percentage-point advantage is somewhat reduced. This is because CGO variants introduce additional calls to Go standard libraries and foreign function interfaces, which Capslock explicitly flags (e.g., via \texttt{CGO\_CAPABILITY}). However, Capslock often raises alerts on the presence of these foreign interfaces rather than on the core malicious logic itself (e.g., actual file writes or exfiltration operations).

\Definition{On the code the application actually executes, GoLeash detects 96--100\% of injections on jointly-visible packages, substantially outperforming Capslock's static analysis. The two techniques fail under largely disjoint circumstances, GoLeash on unexercised code and Capslock on dynamic and fine-grained behavior, so practical defense-in-depth against software supply chain attacks benefits from combining both.}

% -------------------------------------------------------------------------------------------%

\section{Discussion}
\label{sec:discussion}
We here discuss key design choices and considerations that shape the applicability of GoLeash.

\subsection{Workload Coverage}
GoLeash uses dynamic analysis to generate fine-grained, per-package policies based on observed behavior. Its coverage is thus limited by the code exercised during analysis. If (1) rarely executed paths are missed, and (2) those paths use additional capabilities, GoLeash may flag them as violations at runtime (i.e., false positives). This is a known limitation of dynamic analysis \cite{rostamipoor_confine_2023}, not unique to our system. 
Several mitigations apply. First, this problem is mitigated in production-grade software with mature integration test suites, as is the norm for the critical cloud systems we evaluate; integration tests are designed to exercise the dominant control-flow paths and therefore cover most legitimately required capabilities. Second, GoLeash supports incremental refinement of policies: capabilities that trigger false positives in staging or canary deployments can be reviewed and added to the allowlist without discarding prior entries (Section~\ref{sec:policy_management}). Third, dynamic analysis can be augmented with workload generation techniques such as fuzzing to systematically expand coverage prior to deployment. Fourth, a hybrid pipeline that uses static analysis (e.g., Capslock) to identify capabilities reachable in source but not exercised under the workload would let developers preemptively review the gap before deploying GoLeash in enforcement mode.

\subsection{Capability Reuse}
GoLeash enforces policies by detecting when packages invoke unauthorized capabilities. However, if malware is injected into a permissive module and reuses already-allowed capabilities, it may evade detection. 
This piggybacking strategy can bypass capability filters if malicious actions stay within the originally authorized set. Our evaluation, which injects malware into every package, already reflects this, as seen in the observed non-detection rates. A complementary question is how tight inferred policies are in practice, that is, how much residual attack surface remains for an adversary willing to stay within the allowed capabilities of a compromised package. Tightness depends primarily on two factors: (i) the diversity of capabilities a package legitimately exercises, and (ii) the specificity of the call paths recorded for each capability. In our evaluation, the median package in Kubernetes exercises only three capabilities along a small number of distinct call paths (see Section~\ref{sec:evaluation}), which leaves narrow room for piggybacking. Conversely, packages that legitimately invoke broad capabilities offer the attacker more latitude: malicious code injected into such a package could, in principle, exfiltrate data over the same connection paths used for legitimate traffic without triggering a violation. Fully bounding this threat requires an ecosystem-wide characterization of how capabilities are distributed across Go packages, which remains important future work to let policy generators reason quantitatively about per-package piggybacking risk.

\subsection{Fine-Grained Policy Enforcement}
Currently, GoLeash enforces policies at the boundary of discrete capabilities. Moving beyond this binary model to more fine-grained enforcement can further restrict what compromised packages can achieve and address the residual attack surfaces discussed above. First, deepening the enforcement requires inspection of syscall arguments and side effects (e.g., destination IPs, file paths), as proposed in prior work \cite{zhang2020dynamic, rostamipoor_confine_2023, pailoor2020automated}. Such analysis is orthogonal to our focus on per-package capability boundaries but can be integrated into GoLeash. Second, a promising refinement is to enforce not only which capabilities a package may invoke, but also how many times and in what order. For example, packages that read a configuration file once at initialization (a common pattern in TLS libraries) could be constrained to a single \texttt{open}+\texttt{read} sequence, with all subsequent filesystem access denied. The eBPF tracing infrastructure already captures the information needed to implement this (per-package syscall counts and ordered call paths), and integrating finite-state automata over capability sequences is a clear extension path for future work.

\subsection{Portability Across Languages and Systems}
GoLeash is designed for the Go language on Linux, leveraging eBPF tracing and Go-specific features (e.g., package management, symbol tables). However, its core design principles (e.g., stack-based syscall attribution, per-component capability policies), are conceptual and portable to other languages and systems. For instance, similar techniques could apply to Java and Rust, which also support package management and expose symbols and package information in runtime stack traces. Additionally, other operating systems, such as FreeBSD and Windows, also offer production-grade tracing frameworks, such as DTrace \cite{dtrace_on_windows}.

\subsection{Stripped Binaries}
GoLeash currently resolves stack-trace addresses to function names by reading the standard ELF symbol table, which may be omitted from optimized production builds (e.g., by default, Kubernetes ships stripped binaries~\cite{golang_issue}).
In practice, symbols remain accessible to GoLeash in two ways.
First, operators of security-sensitive infrastructure commonly maintain custom build pipelines for hardening and observability purposes; in such pipelines, retaining symbols is a single linker-flag change with a modest binary-size impact, well below the security benefit GoLeash provides. 
A complementary path, which we view as a clear engineering extension, is to adapt GoLeash to parse Go's runtime metadata (\texttt{.gopclntab}) directly. This metadata is preserved across stripping because the Go runtime requires it for panic stack traces, and existing tooling (debug/gosym, GoReSym~\cite{goresym}, redress~\cite{redress}) can already recover function and package names from it. Adopting this path would remove the symbol-table dependency entirely.

% -------------------------------------------------------------------------------------------%
\section{Related Work}
\label{sec:related}
We categorize related work along the five key dimensions from our design: 
need for code changes, 
support for fine-grained access control, 
automated policy generation, support for multi-process apps, 
and detection of obfuscated behavior. We summarize in Table~\ref{tab:comparison-requirements} other systems that are most directly comparable to our one, and discuss at more length the others in the text (e.g., static analysis).

\subsection{Application-Level Sandboxing}
A number of solutions confine or restrict the system-level behavior through runtime sandboxing.  

NatiSand~\cite{abbadini2023natisand} and Cage4Deno~\cite{abbadini2023cage4deno} use Linux security features (e.g., Seccomp, eBPF, Landlock) to restrict native extension in JavaScript runtimes like Node.js and Deno. Natinad relies on manual JSON-based configuration to restrict operations such as file or network access. Cage4Deno adds support for isolating subprocesses, but treats each subprocess as a standalone application, lacking awareness of finer-grained entities like npm packages. Similarly, HODOR~\cite{wang2023hodor} shrinks the attack surface of Node.js applications by auto-generating syscall whitelists. While it achieves thread-level granularity, it cannot distinguish which dependency initiated a syscall, limiting its ability to isolate malicious libraries.

Other efforts focus on container-level hardening. Confine~\cite{rostamipoor_confine_2023} statically generates restrictive Seccomp syscall policies for containers via whole-binary analysis, but lacks intra-container granularity or obfuscation resilience. $\mu$PolicyCraft~\cite{blair2024automated} synthesizes stateful syscall automata for microservices through symbolic execution and enforces them via a telemetry monitor, yet still operats at the microservice/container level rather than inside processes.

Finally, $\mu$SCOPE~\cite{roessler2021muscope} dynamically traces memory accesses and privilege operations in lage codebases (e.g., the Linux kernel) to identify overprivileged regions. While effective for analysis and privilege planning, it does not enforce runtime behavior and lacks syscall-level or per-package enforcement.

Overall, application-wide or ``coarse-level'' sandboxing solutions successfully confine large swaths of code, but they lack per-dependency distinctions, which means that malicious or compromised libraries cannot be individually restricted. In contrast, GoLeash operates at the package level within a single Go binary, with no reliance on manual configuration.

\begin{table*}[t]
    \centering
    \caption{Comparison of Capability Enforcement Approaches Against GoLeash’s Design Requirements}
    \label{tab:comparison-requirements}
    \resizebox{\textwidth}{!}{%
    \begin{tabular}{lccccc}
        \toprule
        \textbf{Approach} & \textbf{No Code} & \textbf{Fine-Grained} & \textbf{Auto} & \textbf{Multi-Process} & \textbf{Detect} \\
         \textbf{} & \textbf{Changes} & \textbf{Control} & \textbf{Policy Gen} & \textbf{Support} & \textbf{Obfuscated Behavior} \\
        \midrule
        MIR~\cite{vasilakis2021preventing} 
            & \xmark & \cmark & \cmark & \xmark & \xmark \\
        & \scriptsize instrumentation at load-time & \scriptsize per-lib + field-level APIs & \scriptsize static + dynamic analysis & \scriptsize single-process JS & \scriptsize JS APIs only \\

        NatiSand~\cite{abbadini2023natisand} 
            & \cmark & \xmark & \cmark & \cmark & \xmark \\
        & \scriptsize external JSON policy only & \scriptsize coarse FS/IPC/NET & \scriptsize trace-based (eBPF/strace) & \scriptsize native libs + subprocesses & \scriptsize not syscall-level \\

        Cage4Deno~\cite{abbadini2023cage4deno} 
            & \cmark & \xmark & \cmark & \xmark & \xmark \\
        & \scriptsize CLI policy, no code mods & \scriptsize file-level RWX only & \scriptsize dynamic tracing tool & \scriptsize Deno subprocesses only & \scriptsize JS runtime only \\

        HODOR~\cite{wang2023hodor} 
            & \cmark & \xmark & \cmark & \xmark & \cmark \\
        & \scriptsize seccomp, no app mods & \scriptsize per-thread, not dep & \scriptsize call graph + tracing & \scriptsize threads only & \scriptsize syscall-level filters \\

        ZTD\textsubscript{JAVA}~\cite{amusuo2023ztd} 
            & \xmark & \cmark & \cmark & \xmark & \cmark \\
        & \scriptsize Java agent + bytecode mod & \scriptsize per-dep + resource-level & \scriptsize runtime tracing & \scriptsize no proc tracking & \scriptsize native/resource-level \\

        $\mu$PolicyCraft~\cite{blair2024automated}
            & \cmark & \xmark & \cmark & \cmark & \xmark \\
        & \scriptsize no code mods, trace-based monitor & \scriptsize microservice/container level & \scriptsize symbolic exec + telemetry & \scriptsize full service orchestration & \scriptsize not syscall-level \\

        Confine~\cite{rostamipoor_confine_2023} 
            & \cmark & \xmark & \cmark & \cmark & \cmark \\
        & \scriptsize no code mods, binary-level & \scriptsize container-level only & \scriptsize static binary analysis & \scriptsize full container, incl. procs & \scriptsize syscall-level \\
 
        \hline
        \textbf{GoLeash (ours)} 
            & \cmark & \cmark & \cmark & \cmark & \cmark \\
        & \scriptsize eBPF, no code/build mods & \scriptsize per-package + syscall-level & \scriptsize call graph + trace & \scriptsize multi-proc + threads & \scriptsize syscall-level enforcement \\
        \bottomrule
    \end{tabular}%
    }
    \label{tab:comparison}
\end{table*}

\subsection{Permissions Systems for Packages}
A second line of work provides package-level or library-level permission systems.

MIR~\cite{vasilakis2021preventing} uses static analysis to assign read-write-execute capabilities to Node.js libraries, applying them at runtime. ZTD\textsubscript{JAVA}~\cite{amusuo2023ztd} similarly enforces package-level permissions in Java via a combination of manual configuration and runtime monitoring. BinWrap~\cite{christou2023binwrap} targets native (binary) Node.js add-ons via a hybrid static analysis of both the JavaScript wrapper and the compiled ELF shared object, automatically inferring permissions and enforcing them at runtime through thread-level isolation and syscall filtering. While these tools reduce risk from vulnerabilities within a dependency, their threat models do not account for maliciously added capabilities, and they rely on static analysis that lacks visibility into dynamically loaded or obfuscated behavior.

Other efforts depend on manually specified permissions or operate over high-level, language-specific API. Ferreira et al.~\cite{ferreira} propose a Node.js permission system configured by developers to constrain packages at the JavaScript API level, with no syscall visibility. FLEXDROID~\cite{seo2016flexdroid} enforces privilege separation in Android apps by intercepting Dalvik calls, but relies on Android permission model and does not observe OS-level behavior.

Other work enforces stronger isolation at the library level using system call interposition. Codejail~\cite{wu2012codejail} transparently confines dynamically linked libraries in Linux via a dual-process model that mediates memory and syscall behavior with the main program. LibVM~\cite{goonasekera2015libvm} builds virtualized execution environments for shared libraries, with hardware-assisted or ptrace-based syscall control. Enclosure~\cite{ghosn2021enclosure} similarly isolates untrusted libraries by associating closures with restricted memory views and syscall filters, enforced via Intel VT-x. It supports package-level policies in Go and Python, but requires developers to annotate code explicitly and integrate language-level constructs. While all three systems support native code and intra-process isolation, they lack automated policy generation, and require developer effort or manual integration into host applications.

Most systems either assume dependencies are benign, require developer effort to specify policies, or lack visibility into system-level behavior. None provide runtime enforcement against malicious capabilities introduced via supply chain compromise. In contrast, GoLeash enforces syscall-level capabilities at the package level, without requiring source changes or developer-defined manifests. Its threat model explicitly includes malicious dependencies and it operates independently of language-specific security models. This makes GoLeash uniquely applicable to Go binaries, which underpin mission-critical software \cite{go_case_studies}, such as Kubernetes.

\subsection{Pre-Deployment Package Vetting}
A significant body of research focuses on mitigating software supply chain attacks via pre-deployment vetting, using static code analysis, dynamic install-time tracing, or a combination of both.

Latch~\cite{wyss2022wolf} dynamically traces system calls during install-script execution of npm packages to infer per-package install-time manifests, and enforces user-defined template policies over those manifests. Ohm et al.~\cite{ohm2023you} propose a differential capability analysis approach that compares new package versions against trusted baselines and restricts capabilities through a modified Node.js runtime. iHunter~\cite{liu2024ihunter} performs static taint analysis on iOS SDKs to detect privacy violations such as cross-library data harvesting, using symbolic execution and NLP-assisted API modeling. While these approaches offer effective pre-deployment vetting, they operate entirely offline and during installation, and do not detect runtime behavior after deployment.

Other tools such as GuardDog and Amalfi~\cite{guarddog, amalfi} scan for known malware patterns, suspicious metadata, or typosquatting during package publication. These techniques help eliminate obviously malicious packages but are blind to capabilities that are invoked dynamically at runtime. 

ARMS~\cite{kalu2025arms} proposes an actor-centric alternative: rather than scanning artifacts, it scores prospective contributors against security-relevant signals (vulnerability remediation history, branch protection, code-scanning adoption) to help maintainers vet pull requests before merge. ARMS operates entirely upstream of execution and addresses a different threat profile than runtime enforcement.

These methods are inherently limited to pre-deployment analysis and cannot respond to runtime malicious behavior. They also rely either on static signatures or on expert-written template policies, which work well when the space of legitimate behaviors is small and well-understood (e.g., npm install scripts) but do not scale to the per-package diversity of runtime behavior in complex multi-package applications. GoLeash instead operates at runtime and captures malicious behavior even if introduced post-installation or obfuscated via native code, with per-package allowlists derived from observed execution. Its runtime enforcement makes it resilient to evasion techniques that bypass pre-deployment vetting.

% -------------------------------------------------------------------------------------------%

\section{Conclusion}
\label{sec:conclusion}
GoLeash is a novel solution for mitigating software supply chain attacks in Go. By enforcing runtime least-privilege boundaries at the level of individual Go packages, GoLeash can detect maliciously added capabilities introduced via compromised dependencies. These fine-grained enforcement mechanisms enable GoLeash to identify stealthy behaviors that application-level dynamic techniques or static analysis miss. Our evaluation shows that GoLeash achieves high detection accuracy, remains effective under obfuscation, and incurs acceptable overhead. As supply chain attacks continue to escalate in sophistication, GoLeash provides a foundation for hardening the runtime security of modern software.

\section*{Acknowledgments}
This work was partially supported by the Swedish Foundation for Strategic Research (SSF), VINNOVA (https://a2i-project.github.io/), and FLEGREA CUP E53D23007950001 (https://flegrea.github.io).

\balance
\bibliographystyle{IEEEtran}
\bibliography{bibfile}

@article{rostamipoor_confine_2023,
    title = {Confine: {Fine}-grained system call filtering for container attack surface reduction},
    volume = {132},
    journal = {Computers \& Security},
    author = {Rostamipoor, Maryam and Ghavamnia, Seyedhamed and Polychronakis, Michalis},
    year = {2023},
}

@inproceedings{christou2023binwrap,
  title={Binwrap: Hybrid protection against native node. js add-ons},
  author={Christou, George and Ntousakis, Grigoris and Lahtinen, Eric and Ioannidis, Sotiris and Kemerlis, Vasileios P and Vasilakis, Nikos},
  booktitle={Proceedings of the 2023 ACM Asia Conference on Computer and Communications Security},
  pages={429--442},
  year={2023}
}

@article{bagherzadeh2018analyzing,
  title={Analyzing a decade of Linux system calls},
  author={Bagherzadeh, Mojtaba and Kahani, Nafiseh and Bezemer, Cor-Paul and Hassan, Ahmed E and Dingel, Juergen and Cordy, James R},
  journal={Empirical Software Engineering},
  volume={23},
  number={3},
  pages={1519--1551},
  year={2018},
  publisher={Springer}
}

@inproceedings{calciati2017apps,
  title={How do apps evolve in their permission requests? a preliminary study},
  author={Calciati, Paolo and Gorla, Alessandra},
  booktitle={2017 IEEE/ACM 14th International Conference on Mining Software Repositories (MSR)},
  pages={37--41},
  year={2017},
  organization={IEEE}
}

@inproceedings{dig2005role,
  title={The role of refactorings in API evolution},
  author={Dig, Danny and Johnson, Ralph},
  booktitle={21st IEEE International Conference on Software Maintenance (ICSM'05)},
  pages={389--398},
  year={2005},
  organization={IEEE}
}

@inproceedings{canella2021automating,
  title={Automating seccomp filter generation for linux applications},
  author={Canella, Claudio and Werner, Mario and Gruss, Daniel and Schwarz, Michael},
  booktitle={Proceedings of the 2021 on Cloud Computing Security Workshop},
  pages={139--151},
  year={2021}
}

@inproceedings{demarinis2020sysfilter,
  title={Sysfilter: Automated system call filtering for commodity software},
  author={DeMarinis, Nicholas and Williams-King, Kent and Jin, Di and Fonseca, Rodrigo and Kemerlis, Vasileios P},
  booktitle={23rd International Symposium on Research in Attacks, Intrusions and Defenses (RAID 2020)},
  pages={459--474},
  year={2020}
}

@article{ohm2023you,
  title={{You Can Run But You Can't Hide: Runtime Protection Against Malicious Package Updates For Node.js}},
  author={Ohm, Marc and Pohl, Timo and Boes, Felix},
  journal={arXiv preprint arXiv:2305.19760},
  year={2023}
}

@inproceedings{wyss2022wolf,
  title={{Wolf at the door: Preventing install-time attacks in npm with latch}},
  author={Wyss, Elizabeth and Wittman, Alexander and Davidson, Drew and De Carli, Lorenzo},
  booktitle={ACM ASIA Conference on Computer and Communications Security (ASIACCS)},
  pages={1139--1153},
  year={2022}
}

@inproceedings{vasilakis2021preventing,
  title={{Preventing Dynamic Library Compromise on Node.js via RWX-Based Privilege Reduction}},
  author={Vasilakis, Nikos and Staicu, Cristian-Alexandru and Ntousakis, Grigoris and Kallas, Konstantinos and Karel, Ben and DeHon, Andr{\'e} and Pradel, Michael},
  booktitle={ACM SIGSAC Conference on Computer and Communications Security (CCS)},
  pages={1821--1838},
  year={2021}
}

@inproceedings{abbadini2023natisand,
  title={{NatiSand: Native code sandboxing for JavaScript runtimes}},
  author={Abbadini, Marco and Facchinetti, Dario and Oldani, Gianluca and Rossi, Matthew and Paraboschi, Stefano},
  booktitle={26th International Symposium on Research in Attacks, Intrusions and Defenses (RAID)},
  pages={639--653},
  year={2023}
}

@inproceedings{abbadini2023cage4deno,
  title={{Cage4Deno: A fine-grained sandbox for Deno subprocesses}},
  author={Abbadini, Marco and Facchinetti, Dario and Oldani, Gianluca and Rossi, Matthew and Paraboschi, Stefano},
  booktitle={ACM ASIA Conference on Computer and Communications Security (ASIACCS)},
  pages={149--162},
  year={2023}
}

@inproceedings{wang2023hodor,
  title={{HODOR: Shrinking Attack Surface on Node.js via System Call Limitation}},
  author={Wang, Wenya and Lin, Xingwei and Wang, Jingyi and Gao, Wang and Gu, Dawu and Lv, Wei and Wang, Jiashui},
  booktitle={ACM SIGSAC Conference on Computer and Communications Security (CCS)},
  pages={2800--2814},
  year={2023}
}

@article{amusuo2023ztd,
  title={ZTD$_{JAVA}$: Mitigating Software Supply Chain Vulnerabilities via Zero-Trust Dependencies},
  author={Amusuo, Paschal C and Robinson, Kyle A and Singla, Tanmay and Peng, Huiyun and Machiry, Aravind and Torres-Arias, Santiago and Simon, Laurent and Davis, James C},
  journal={IEEE/ACM 47th International Conference on Software Engineering (ICSE)},
  year={2025}
}

@misc{golang_issue,
  title = {{generate external DWARD debuginfo archieves directly}},
  author={Github issue},
  howpublished = {\url{https://github.com/golang/go/issues/51692}},
  year = {2022}
}

@misc{etcd_benchmark,
  title = {{etcd - benchmark}},
  author={Github repo},
  howpublished = {\url{https://github.com/etcd-io/etcd/blob/main/tools/benchmark/README.md}},
  year = {2024}
}

@misc{wrk,
  title = {{wg/wrk: Modern HTTP Benchmarking tool}},
  author={Github repo},
  howpublished = {\url{https://github.com/wg/wrk}},
  year = {2015}
}

@misc{go_cacm,
  title = {{Giving Go a Go: Simplifying Cloud Infrastructure Development}},
  author={Alex Williams},
  howpublished = {\url{https://cacm.acm.org/blogcacm/giving-go-a-go-simplifying-cloud-infrastructure-development/}},
  year = {2024}
}

@misc{go_supply_chain_attack_boltdb,
  title = {{Go Supply Chain Attack: Malicious Package Exploits Go Module Proxy Caching for Persistence}},
  author = {Kirill Boychenko},
  howpublished = {\url{https://socket.dev/blog/malicious-package-exploits-go-module-proxy-caching-for-persistence}},
  year = {2025}
}

@misc{hypert,
  title = {{Typosquatted Go Packages Deliver Malware Loader Targeting Linux and macOS Systems}},
  author = {Kirill Boychenko},
  howpublished = {\url{https://socket.dev/blog/typosquatted-go-packages-deliver-malware-loader}},
  year = {2025}
}

@misc{go_case_studies,
  title = {{Case Studies}},
  author = {Go Dev},
  howpublished = {\url{https://go.dev/solutions/case-studies}},
  year = {2025}
}

@misc{linux_tracepoints,
    author = {{The Linux Kernel Archives}},
    title = {{Using the Linux Kernel Tracepoints}},
    howpublished = {\url{https://docs.kernel.org/trace/tracepoints.html}},
    year = {2025}
}

@misc{repojacking_go,
    title = {{Hijackable Go Module Repositories}},
    author = {Jacob Baines},
    howpublished = {\url{https://vulncheck.com/blog/go-repojacking}},
    year = {2023}
}

@misc{capslock,
    title = {{Google Capslock}},
    author = {{Google Inc.}},
    howpublished = {\url{https://github.com/google/capslock}},
    year = {2025}
}

@article{smith2012contemporary,
  title={{A contemporary look at Saltzer and Schroeder's 1975 design principles}},
  author={Smith, Richard E},
  journal={IEEE Security \& Privacy},
  volume={10},
  number={6},
  pages={20--25},
  year={2012},
  publisher={IEEE}
}

@article{saltzer1975protection,
  title={The protection of information in computer systems},
  author={Saltzer, Jerome H and Schroeder, Michael D},
  journal={Proceedings of the IEEE},
  volume={63},
  number={9},
  pages={1278--1308},
  year={1975},
  publisher={IEEE}
}

@inproceedings{cesarano2023gosurf,
  title={GoSurf: Identifying Software Supply Chain Attack Vectors in Go},
  author={Cesarano, Carmine and Andersson, Vivi and Natella, Roberto and Monperrus, Martin},
  booktitle={Proceedings of the 2024 Workshop on Software Supply Chain Offensive Research and Ecosystem Defenses},
  pages={33--42},
  year={2023}
}

@inproceedings{wan2017mining,
  title={Mining sandboxes for linux containers},
  author={Wan, Zhiyuan and Lo, David and Xia, Xin and Cai, Liang and Li, Shanping},
  booktitle={2017 IEEE International Conference on Software Testing, Verification and Validation (ICST)},
  pages={92--102},
  year={2017},
  organization={IEEE}
}

@article{wan2019practical,
  title={Practical and effective sandboxing for Linux containers},
  author={Wan, Zhiyuan and Lo, David and Xia, Xin and Cai, Liang},
  journal={Empirical Software Engineering},
  volume={24},
  number={6},
  pages={4034--4070},
  year={2019},
  publisher={Springer}
}

@inproceedings{comparetti2010identifying,
  title={Identifying dormant functionality in malware programs},
  author={Comparetti, Paolo Milani and Salvaneschi, Guido and Kirda, Engin and Kolbitsch, Clemens and Kruegel, Christopher and Zanero, Stefano},
  booktitle={2010 IEEE Symposium on Security and Privacy},
  pages={61--76},
  year={2010},
  organization={IEEE}
}

@article{williams2025research,
author = {Williams, Laurie and Benedetti, Giacomo and Hamer, Sivana and Paramitha, Ranindya and Rahman, Imranur and Tamanna, Mahzabin and Tystahl, Greg and Zahan, Nusrat and Morrison, Patrick and Acar, Yasemin and Cukier, Michel and K\"{a}stner, Christian and Kapravelos, Alexandros and Wermke, Dominik and Enck, William},
title = {{Research Directions in Software Supply Chain Security}},
year = {2025},
journal = {ACM Trans. Softw. Eng. Methodol.}
}

@misc{kubernetes,
  title = {Kubernetes: Production-Grade Container Orchestration},
  author = {{Kubernetes Authors}},
  year = {2024},
  howpublished = {\url{https://kubernetes.io}},
  note = {Accessed: 2025-04-11}
}

@misc{docker,
  title = {Docker: Empowering App Development for Developers},
  author = {{Docker Inc.}},
  year = {2024},
  howpublished = {\url{https://www.docker.com}},
  note = {Accessed: 2025-04-11}
}

@misc{terraform,
  title = {Terraform by HashiCorp},
  author = {{HashiCorp}},
  year = {2024},
  howpublished = {\url{https://www.terraform.io}},
  note = {Accessed: 2025-04-11}
}

@misc{etcd,
  title = {etcd: A Distributed, Reliable Key-Value Store for the Most Critical Data of a Distributed System},
  author = {{etcd Authors}},
  year = {2024},
  howpublished = {\url{https://etcd.io}},
  note = {Accessed: 2025-04-11}
}

@misc{cisa_ssc,
  author = {{CISA and NSA}},
  title = {Defending Against Software Supply Chain Attacks},
  year = {2021},
  howpublished = {\url{https://www.cisa.gov/sites/default/files/publications/defending_against_software_supply_chain_attacks_508.pdf}},
  note = {Accessed: 2025-04-11}
}

@misc{eventstream_attack,
  author = {{GitHub Community}},
  title = {Malicious code found in npm package event-stream},
  year = {2018},
  howpublished = {\url{https://github.com/dominictarr/event-stream/issues/116}},
  note = {Accessed: 2025-04-11}
}

@misc{ebpf_foundation,
  author = {{eBPF Foundation}},
  title = {What is eBPF?},
  year = {2024},
  howpublished = {\url{https://ebpf.io/what-is-ebpf/}},
  note = {Accessed: 2025-04-11}
}

@misc{gvisor,
  author = {{gVisor Team}},
  title = {gVisor: The Container Security Platform},
  year = {2024},
  howpublished = {\url{https://gvisor.dev/}},
  note = {Accessed: 2025-04-11}
}

@misc{seccomp_linux,
  author = {{The Linux Kernel Community}},
  title = {seccomp(2) — Linux manual page},
  year = {2024},
  howpublished = {\url{https://man7.org/linux/man-pages/man2/seccomp.2.html}},
  note = {Accessed: 2025-04-11}
}

@misc{go-modules-blog,
  author       = {Tyler Bui},
  title        = {Using Go Modules - The Go Programming Language},
  year         = {2019},
  url          = {https://blog.golang.org/using-go-modules},
  note         = {Accessed: 2025-04-12}
}

@misc{go-runtime-docs,
  title        = {Go standard library - the runtime package},
  author       = {{Go Team}},
  year         = {2023},
  url          = {https://pkg.go.dev/runtime},
  note         = {Accessed April 2025}
}

@misc{go-stdlib,
  title        = {Go standard library - Go Packages},
  author       = {{Go Team}},
  year         = {2023},
  url          = {https://pkg.go.dev/std},
  note         = {Accessed April 2025}
}

@misc{depsdev,
  author       = {{Open Source Insights Team}},
  title        = {Open Source Insights},
  year         = {2025},
  url          = {https://deps.dev/},
  note         = {Accessed April 12, 2025}
}

@misc{github,
  author       = {{GitHub}},
  title        = {GitHub},
  year         = {2025},
  url          = {https://github.com},
  note         = {Accessed April 12, 2025}
}

@misc{isovalent,
  author       = {{Isovalent}},
  title        = {eBPF-based networking, security, and observability},
  year         = {2025},
  url          = {https://isovalent.com/},
  note         = {Accessed: 2025-04-12}
}

@misc{tetragon,
  author       = {{Tetragon}},
  title        = {eBPF-based Security Observability and Runtime Enforcement},
  year         = {2025},
  url          = {https://tetragon.io/},
  note         = {Accessed: 2025-04-12}
}

@misc{cilium,
  author       = {{Cilium}},
  title        = {eBPF-based Networking, Observability, Security},
  year         = {2025},
  url          = {https://cilium.io/},
  note         = {Accessed: 2025-04-12}
}

@misc{falco,
  author       = {{Falco}},
  title        = {Detect security threats in real time},
  year         = {2025},
  url          = {https://falco.org/},
  note         = {Accessed: 2025-04-12}
}

@article{molina2025light,
  title={Light up that Droid! On the effectiveness of static analysis features against app obfuscation for Android malware detection},
  author={Molina-Coronado, Borja and Ruggia, Antonio and Mori, Usue and Merlo, Alessio and Mendiburu, Alexander and Miguel-Alonso, Jose},
  journal={Journal of Network and Computer Applications},
  volume={235},
  pages={104094},
  year={2025},
  publisher={Elsevier}
}

@article{khoury2012security,
  title={Which security policies are enforceable by runtime monitors? a survey},
  author={Khoury, Rapha{\"e}l and Tawbi, Nadia},
  journal={Computer Science Review},
  volume={6},
  number={1},
  pages={27--45},
  year={2012},
  publisher={Elsevier}
}

@inproceedings{siadati2024devphish,
  title={{DevPhish: Exploring Social Engineering in Software Supply Chain Attacks on Developers}},
  author={Siadati, Hossein and Jafarikhah, Sima and Sahin, Elif and Hernandez, Terrence and Tripp, Elijah and Khryashchev, Denis and Kharraz, Amin},
  booktitle={2024 IEEE 15th Annual Ubiquitous Computing, Electronics \& Mobile Communication Conference (UEMCON)},
  pages={517--523},
  year={2024},
  organization={IEEE}
}

@article{cox2025fifty,
  title={{Fifty Years of Open Source Software Supply Chain Security}},
  author={Cox, Russ},
  journal={ACM Queue},
  volume={23},
  number={1},
  pages={84--107},
  year={2025}
}

@misc{bpf2go,
  author       = {{cilium}},
  title        = {bpf2go},
  year         = {2025},
  url          = {https://github.com/cilium/ebpf/tree/main/cmd/bpf2go},
  note         = {Accessed: 2025-04-12}
}

@misc{cwe441,
  author       = {{MITRE}},
  title        = {CWE-441: Unintended Proxy or Intermediary ('Confused Deputy')},
  year         = {2025},
  url          = {https://cwe.mitre.org/data/definitions/441.html},
  note         = {Accessed: 2025-04-12}
}

@misc{linux_syscall_table,
  author       = {{The Linux Kernel Community}},
  title        = {x86\_64 System Call Table (syscall\_64.tbl)},
  year         = {2025},
  url          = {https://github.com/torvalds/linux/blob/v6.7/arch/x86/entry/syscalls/syscall_64.tbl},
  note         = {Accessed: 2025-04-12}
}

@misc{coredns,
  author       = {{CoreDNS Authors}},
  title        = {CoreDNS: DNS Server That Chains Plugins},
  year         = {2025},
  url          = {https://github.com/coredns/coredns},
  note         = {Accessed: 2025-04-12}
}

@misc{frp,
  author       = {{fatedier}},
  title        = {frp: A Fast Reverse Proxy to Help You Expose a Local Server Behind a NAT or Firewall to the Internet},
  year         = {2025},
  url          = {https://github.com/fatedier/frp},
  note         = {Accessed: 2025-04-12}
}

@misc{goethereum,
  author       = {{Ethereum Foundation}},
  title        = {go-ethereum: Go Implementation of the Ethereum Protocol},
  year         = {2025},
  url          = {https://github.com/ethereum/go-ethereum},
  note         = {Accessed: 2025-04-12}
}

@article{o2011obfuscation,
  title={Obfuscation: The hidden malware},
  author={O'Kane, Philip and Sezer, Sakir and McLaughlin, Kieran},
  journal={IEEE Security \& Privacy},
  volume={9},
  number={5},
  pages={41--47},
  year={2011},
  publisher={Ieee}
}

@inproceedings{xu2012power,
  title={The power of obfuscation techniques in malicious JavaScript code: A measurement study},
  author={Xu, Wei and Zhang, Fangfang and Zhu, Sencun},
  booktitle={2012 7th International Conference on Malicious and Unwanted Software},
  pages={9--16},
  year={2012},
  organization={IEEE}
}

@inproceedings{seo2016flexdroid,
  title={FLEXDROID: Enforcing In-App Privilege Separation in Android.},
  author={Seo, Jaebaek and Kim, Daehyeok and Cho, Donghyun and Shin, Insik and Kim, Taesoo},
  booktitle={NDSS},
  year={2016}
}

@misc{selinux,
  author       = {{SELinux}},
  title        = {SELinuxProject},
  year         = {2025},
  url          = {https://github.com/SELinuxProject/selinux},
  note         = {Accessed: 2025-04-12}
}

@misc{apparmor,
  author       = {{AppArmor}},
  title        = {AppArmor},
  year         = {2025},
  url          = {https://apparmor.net/},
  note         = {Accessed: 2025-04-12}
}

@inproceedings{ferreira,
author = {Ferreira, Gabriel and Jia, Limin and Sunshine, Joshua and K\"{a}stner, Christian},
title = {Containing Malicious Package Updates in npm with a Lightweight Permission System},
year = {2021},
publisher = {IEEE Press},
url = {https://doi.org/10.1109/ICSE43902.2021.00121},
booktitle = {Proceedings of the 43rd International Conference on Software Engineering},
pages = {1334–1346},
numpages = {13},
series = {ICSE '21}
}

@inproceedings{amalfi,
author = {Sejfia, Adriana and Sch\"{a}fer, Max},
title = {Practical automated detection of malicious npm packages},
year = {2022},
isbn = {9781450392211},
publisher = {Association for Computing Machinery},
url = {https://doi.org/10.1145/3510003.3510104},
doi = {10.1145/3510003.3510104},
booktitle = {Proceedings of the 44th International Conference on Software Engineering},
pages = {1681–1692},
numpages = {12},
series = {ICSE '22}
}

@misc{guarddog,
  author       = {{Datadog}},
  title        = {GuardDog: CLI Tool to Identify Malicious PyPI, npm Packages},
  year         = {2024},
  url          = {https://github.com/DataDog/guarddog},
  note         = {Accessed: 2025-04-12}
}

@inproceedings{ohm2020backstabber,
  title={Backstabber’s knife collection: A review of open source software supply chain attacks},
  author={Ohm, Marc and Plate, Henrik and Sykosch, Arnold and Meier, Michael},
  booktitle={Detection of Intrusions and Malware, and Vulnerability Assessment: 17th International Conference, DIMVA 2020, Lisbon, Portugal, June 24--26, 2020, Proceedings 17},
  pages={23--43},
  year={2020},
  organization={Springer}
}

@misc{datadog_malicious_dataset,
  author       = {{Datadog}},
  title        = {Malicious Software Packages Dataset},
  year         = {2025},
  url          = {https://github.com/DataDog/malicious-software-packages-dataset},
  note         = {Accessed: 2025-04-12}
}

@article{kalu2025arms,
  title={ARMS: A Vision for Actor Reputation Metric Systems in the Open-Source Software Supply Chain},
  author={Kalu, Kelechi G and Okorafor, Sofia and Durak, Bet{\"u}l and Laine, Kim and Moreno, Radames C and Torres-Arias, Santiago and Davis, James C},
  journal={arXiv preprint arXiv:2505.18760},
  year={2025}
}

@misc{dtrace_on_windows,
  author       = {{Microsoft}},
  title        = {DTrace on Windows},
  year         = {2025},
  url          = {https://github.com/microsoft/DTrace-on-Windows},
  note         = {Accessed: 2025-04-12}
}

@inproceedings{zhang2020dynamic,
  title={Dynamic malware analysis with feature engineering and feature learning},
  author={Zhang, Zhaoqi and Qi, Panpan and Wang, Wei},
  booktitle={Proceedings of the AAAI conference on artificial intelligence},
  volume={34},
  number={01},
  pages={1210--1217},
  year={2020}
}

@article{pailoor2020automated,
  title={Automated policy synthesis for system call sandboxing},
  author={Pailoor, Shankara and Wang, Xinyu and Shacham, Hovav and Dillig, Isil},
  journal={Proceedings of the ACM on Programming Languages},
  volume={4},
  number={OOPSLA},
  pages={1--26},
  year={2020},
  publisher={ACM New York, NY, USA}
}

@misc{goresym,
  author       = {goresym},
  title        = {Go symbol recovery tool},
  year         = {2025},
  url          = {https://github.com/mandiant/GoReSym},
  note         = {Accessed: 2025-04-12}
}

@misc{redress,
  author       = {goretk},
  title        = {Redress: A Tool for Analyzing Stripped Go Binaries},
  year         = {2025},
  url          = {https://github.com/goretk/redress},
  note         = {Accessed: 2025-04-12}
}

@misc{dnsperf,
  author       = {{DNS-OARC}},
  title        = {dnsperf: DNS Performance Testing Tools},
  year         = {2025},
  url          = {https://github.com/DNS-OARC/dnsperf},
  note         = {Accessed: 2025-04-12}
}

@inproceedings{ghosn2021enclosure,
  title={Enclosure: language-based restriction of untrusted libraries},
  author={Ghosn, Adrien and Kogias, Marios and Payer, Mathias and Larus, James R and Bugnion, Edouard},
  booktitle={Proceedings of the 26th ACM International Conference on Architectural Support for Programming Languages and Operating Systems},
  pages={255--267},
  year={2021}
}

@inproceedings{blair2024automated,
  title={Automated Synthesis of Effect Graph Policies for Microservice-Aware Stateful System Call Specialization},
  author={Blair, William and Araujo, Frederico and Taylor, Teryl and Jang, Jiyong},
  booktitle={2024 IEEE Symposium on Security and Privacy (SP)},
  pages={4554--4572},
  year={2024},
  organization={IEEE}
}

@inproceedings{wu2012codejail,
  title={Codejail: Application-transparent isolation of libraries with tight program interactions},
  author={Wu, Yongzheng and Sathyanarayan, Sai and Yap, Roland HC and Liang, Zhenkai},
  booktitle={Computer Security--ESORICS 2012: 17th European Symposium on Research in Computer Security, Pisa, Italy, September 10-12, 2012. Proceedings 17},
  pages={859--876},
  year={2012},
  organization={Springer}
}

@inproceedings{liu2024ihunter,
  title={$\{$iHunter$\}$: Hunting Privacy Violations at Scale in the Software Supply Chain on $\{$iOS$\}$},
  author={Liu, Dexin and Xiao, Yue and Zhang, Chaoqi and Xie, Kaitao and Bai, Xiaolong and Zhang, Shikun and Xing, Luyi},
  booktitle={33rd USENIX Security Symposium (USENIX Security 24)},
  pages={5663--5680},
  year={2024}
}

@article{goonasekera2015libvm,
  title={LibVM: an architecture for shared library sandboxing},
  author={Goonasekera, Nuwan and Caelli, William and Fidge, Colin},
  journal={Software: Practice and Experience},
  volume={45},
  number={12},
  pages={1597--1617},
  year={2015},
  publisher={Wiley Online Library}
}

@inproceedings{roessler2021muscope,
  title={Mscope: A methodology for analyzing least-privilege compartmentalization in large software artifacts},
  author={Roessler, Nick and Atayde, Lucas and Palmer, Imani and McKee, Derrick and Pandey, Jai and Kemerlis, Vasileios P and Payer, Mathias and Bates, Adam and Smith, Jonathan M and DeHon, Andre and others},
  booktitle={Proceedings of the 24th International Symposium on Research in Attacks, Intrusions and Defenses},
  pages={296--311},
  year={2021}
}

@inproceedings{craun2024eliminating,
  title={Eliminating eBPF Tracing Overhead on Untraced Processes},
  author={Craun, Milo and Hussain, Khizar and Gautam, Uddhav and Ji, Zhengjie and Rao, Tanuj and Williams, Dan},
  booktitle={Proceedings of the ACM SIGCOMM 2024 Workshop on eBPF and Kernel Extensions},
  pages={16--22},
  year={2024}
}

@misc{goleash_artifact,
  author       = {Anonymous Author(s)},
  title        = {{Replication Package for "GoLeash: Mitigating GoLang Software Supply Chain Attack with Runtime Polocy Enforcement"}},
  howpublished = {\url{https://figshare.com/s/c4844c75354df7e94b55}},
  year         = {2025},
  note         = {Accessed: 2025-04-15}
}

@misc{go_evil_packages,
  author       = {Michen Riksen},
  title        = {{Finding Evil Go Packages}},
  howpublished = {\url{https://michenriksen.com/archive/blog/finding-evil-go-packages/}},
  year         = {2025},
  note         = {Accessed: 2025-04-15}
}

@misc{phylum_npm_reverse_2023,
  author = {{The Hacker News}},
  title = {48 Malicious npm Packages Found Deploying Reverse Shells on Developer Systems},
  year = {2023},
  month = {November},
  howpublished = {\url{https://thehackernews.com/2023/11/48-malicious-npm-packages-found.html}},
  note = {Accessed: 2025-12}
}

@misc{pytorch_torchtriton_2022,
  author = {{PyTorch Team}},
  title = {Compromised {PyTorch}-nightly dependency chain between {December} 25th and 30th, 2022},
  year = {2022},
  month = {December},
  howpublished = {\url{https://pytorch.org/blog/compromised-nightly-dependency/}},
  note = {Accessed: 2025-12}
}

@misc{mandiant_uaparser_2021,
  author = {{Mandiant Managed Defense}},
  title = {No unaccompanied miners: supply chain compromises through {N}ode.js packages},
  year = {2021},
  month = {November},
  howpublished = {\url{https://cloud.google.com/blog/topics/threat-intelligence/supply-chain-node-js}},
  note = {Accessed: 2025-12}
}

@misc{zscaler_termncolor_2025,
  author = {Prajapati, Manisha Ramcharan and Singh, Satyam},
  title = {Supply Chain Risk in Python: Termncolor and Colorinal Explained},
  year = {2025},
  month = {August},
  howpublished = {\url{https://www.zscaler.com/blogs/security-research/supply-chain-risk-python-termncolor-and-colorinal-explained}},
  note = {Accessed: 2025-12}
}

@misc{sonatype_klow_2021,
  author = {ElShakankiry, Ali},
  title = {Newly found npm malware mines cryptocurrency on {W}indows, {L}inux, {macOS} devices},
  year = {2021},
  month = {October},
  howpublished = {\url{https://www.sonatype.com/blog/newly-found-npm-malware-mines-cryptocurrency-on-windows-linux-macos-devices}},
  note = {Accessed: 2025-12}
}

\vfill

\end{document}